\documentclass[aps,prl,reprint, superscriptaddress]{revtex4-2}

\usepackage{amsmath}
\usepackage{amssymb}
\usepackage{amsthm}
\usepackage{graphicx}
\usepackage{stmaryrd}
\usepackage{siunitx}
\usepackage{ulem}
\usepackage[usenames,dvipsnames]{xcolor}

\theoremstyle{plain}

\usepackage{mathtools}
\usepackage{pgfplots}
\pgfplotsset{/pgf/number format/use comma,compat=newest}

\newcommand\upd{\mathrm{d}}

\renewcommand\epsilon{\varepsilon}

\usepackage[hidelinks]{hyperref}
\begin{document}


\title{
Flattened and wrinkled encapsulated droplets: Shape-morphing induced by gravity and evaporation}

\author{Davide Riccobelli}
\thanks{These two authors contributed equally.}
\affiliation{MOX -- Dipartimento di Matematica, Politecnico di Milano, Italy}
\author{Hedar H. Al-Terke}
\thanks{These two authors contributed equally.}
\affiliation{Department of Applied Physics, Aalto University School of Science, Finland}
\affiliation{Center of Excellence in Life--Inspired Hybrid Materials (LIBER), Aalto University, Finland}
\author{P\"{a}ivi Laaksonen}
\affiliation{HAMK Tech,  Häme University of Applied Sciences}
\author{Pierangelo Metrangolo}
\affiliation{Department of Chemistry, Materials, and Chemical Engineering “Giulio Natta”, Politecnico di Milano, Italy}
\affiliation{Department of Applied Physics, Aalto University School of Science, Finland}
\affiliation{Center of Excellence in Life--Inspired Hybrid Materials (LIBER), Aalto University, Finland}
\author{Arja Paananen}
\affiliation{VTT Technical Research Centre of Finland Ltd, Finland}
\author{Robin H. A. Ras}
\affiliation{Department of Applied Physics, Aalto University School of Science, Finland}
\affiliation{Center of Excellence in Life--Inspired Hybrid Materials (LIBER), Aalto University, Finland}
\author{Pasquale Ciarletta}
\affiliation{MOX -- Dipartimento di Matematica, Politecnico di Milano, Italy}
\author{Dominic Vella}
\email{dominic.vella@maths.ox.ac.uk}
\affiliation{Mathematical Institute, University of Oxford, Woodstock Rd, Oxford, OX2 6GG, UK.}


\date{\today}

\begin{abstract}
We report surprising morphological changes of suspension droplets (containing class II hydrophobin protein HFBI from \emph{Trichoderma reesei} in water) as they evaporate with a  contact line pinned  on a rigid solid substrate. Both pendant and sessile droplets display the formation of an encapsulating elastic film as the bulk concentration of solute  reaches a critical value during evaporation, but the morphology of the droplet varies significantly: for sessile droplets, the elastic film ultimately crumples in a nearly flattened area close to the apex while in pendant droplets, circumferential wrinkling occurs close to the contact line. These different morphologies are  understood through a gravito-elasto-capillary model that predicts the droplet morphology and the onset of shape changes, as well as showing that the influence of the direction of gravity remains crucial even for very small droplets (where the effect of gravity can normally be neglected). The results pave the way to control droplet shape in  several engineering and biomedical applications.
\end{abstract}

\maketitle

 The shape of a liquid drop resting on a rigid solid surface is governed by a balance between capillary forces and  hydrostatic pressure. For small water droplets, capillarity dominates this balance and the droplet adopts the shape of a spherical cap --- a label its shape retains even if it subsequently  evaporates, albeit with changing contact angle or radius \cite{de2004capillarity}. While larger water drops may be influenced by gravity, and hence start with more exotic shapes, upon evaporation they must ultimately reach the small droplet limit, and hence adopt the spherical cap shape too. This simple picture of droplets always evaporating as spherical caps does not, however, hold for the evaporation of more complex 
 droplets such as mixtures of water and solid particles or emulsions. In particular, the evaporation of such drops may lead to non-spherical shapes \cite{Bala_Subramaniam_2005,Abe_2017,Kaufman_2017, Kim_2019}, reflecting the non-trivial behavior of interfaces including solid-like films formed on the surface from the aggregation of contaminants \cite{Subramaniam_2005,Yamasaki_2015,Yamasaki_2016,Abe_2017,Kaufman_2017, Kim_2019}. In these settings and others \cite{Denkov2015,Haas2017,Giomi2021,Peddireddy2021} the non-trivial changes in shape that occur are controlled by various balances between surface tension and elasticity and are not generally sensitive to body forces such as gravity.

At the same time, the creation of a capsule around a droplet 
is useful to protect the inner material from 
the external environment and, possibly, to selectively release the droplet's contents under specific conditions \cite{Dinsmore2002}.
Encapsulation is achieved using a variety of methods \cite{Bah_2020}, including \emph{in-situ} polymerization \cite{Brown_2003}, self-assembly of the capsule \cite{brinker1999evaporation}, elasto-capillary interactions \cite{Py_2007}, spray drying \cite{I_R__1998}, and gravity-induced encapsulation \cite{Abkarian_2013}.

\begin{figure}[t!]
\includegraphics[width=\columnwidth]{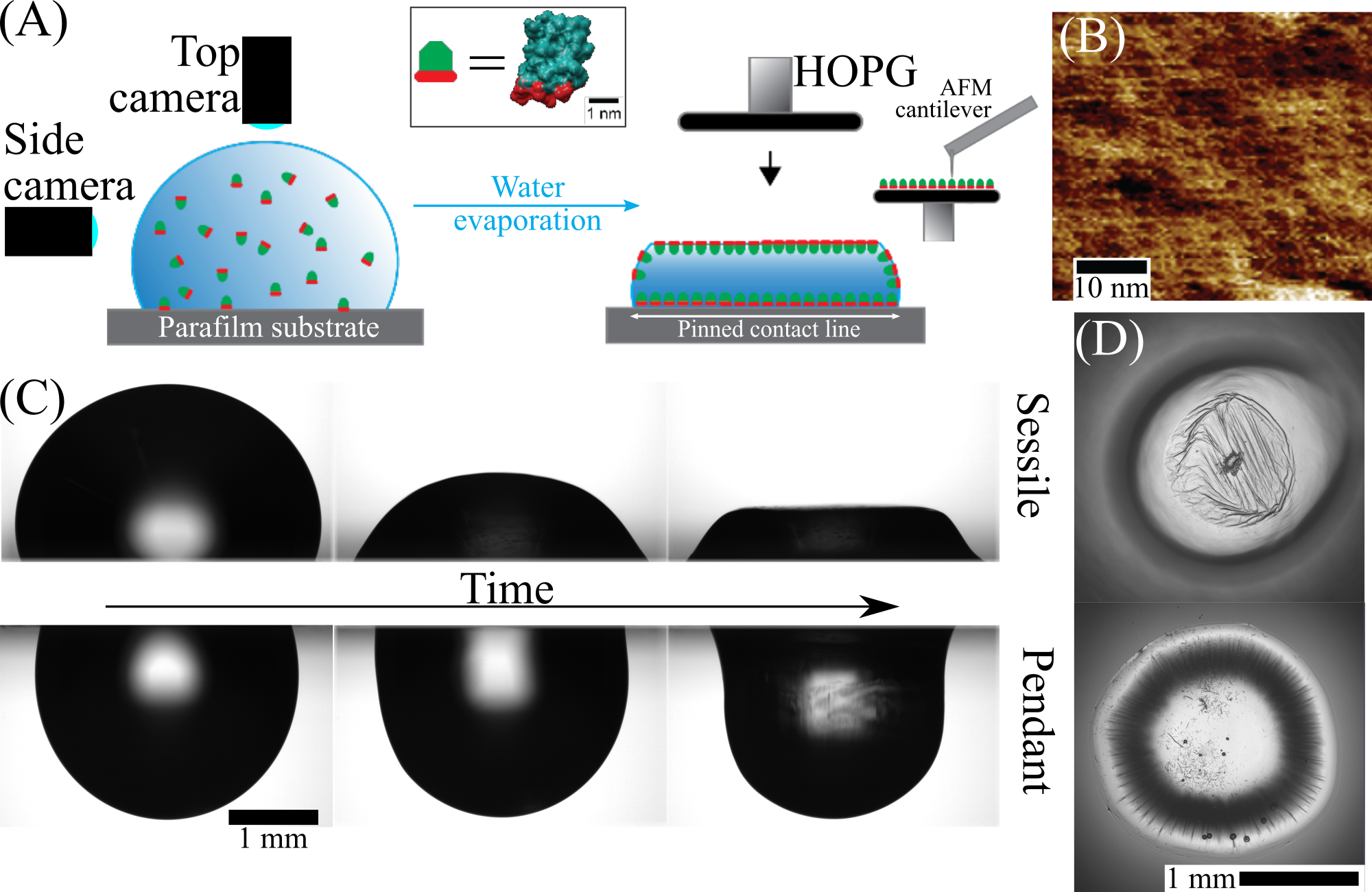}
  \caption{(A): Sketch of the experimental setup. A droplet  is placed on a Parafilm substrate and its shape is recorded from above and the side by cameras. After the formation of the flattened region, a highly ordered pyrolytic graphite (HOPG) substrate was brought in contact with the top of the droplet to perform imaging through an atomic force microscopy (AFM).   (B): AFM image of HFBI monolayer at the air-water surface. (C):  Sessile (top) and pendant (bottom) droplet shapes at different times during evaporation. The initial volume is $V_i=\SI{13.1}{\milli\meter^3}$ in both cases, while the initial concentration of HFBI $\SI{2}{\micro\mole\per\liter}$ for the sessile drop and $\SI{4}{\micro\mole\per\liter}$ for the pendant drop. (D): Plan views of a crumpled sessile (left) and a wrinkled pendant (right) droplet.
}
  \label{fig:1}
\end{figure}

Perhaps the simplest method of capsule formation involves a surface-active molecule spontaneously adsorbing at an interface. Here, the properties of the capsule depend on the interfacial concentration of molecules: at low concentrations the interface has a simple interfacial tension but at higher concentrations also develops a solid-like shear modulus \cite{Bos2001}. In this Letter, we show that the combination of concentration-dependence and external fields (specifically gravity) gives rise to morphological changes in hydrophobin--water droplets, on a solid substrate, as the water phase evaporates.  Specifically, we use class II hydrophobin protein HFBI from  \emph{Trichoderma reesei}.
 HFBI is a water soluble protein and yet is very amphiphilic, readily assembling at  air--water or oil--water interfaces \cite{hahl2019dynamic}.
As shown in FIG.~\ref{fig:1}(C), a sessile droplet (supported against gravity by the surface) starts off as a cap-shape but ultimately forms a flat region close to its apex.  Conversely, a pendant droplet (which hangs beneath the surface under gravity) forms wrinkles close to the contact line. 
While the edge-wrinkling of pendant drops has been modelled previously as an elasto-capillary instability  \cite{Knoche_2013}, the formation of a central flat spot in the sessile case is still debated: it has previously been proposed that the buoyancy of HFBI molecules drives them to form a raft that then floats to the top (sessile case) or sinks to the edge (pendant case) and forces the interface to be flat in these regions  \cite{Yamasaki_2015,Yamasaki_2016}. While these experimental observations (and others on a slope in which the normal to the flat region is parallel to the direction of gravity) strongly suggest that gravity plays a key role in the phenomenon, it stretches credulity that individual molecules (albeit large ones) are affected by gravity to this extent ---  a simple calculation shows \footnote{Please see the Supplemental Material for details on the experimental setup and the mathematical model.} that the sedimentation length \cite{Piazza2012} for these molecules $\ell_s\approx200\mathrm{~m}$.
Given this gravitational height is so much larger than any length scale in the experiment, we seek to understand how gravity enters the problem, focusing on the role played by its effect on droplet shape. We begin by outlining our own experimental procedures.

Water droplets, with initial volume $V_i$ in the range $\SI{8.63}-\SI{18.68}{\milli\meter^3}$,  containing HFBI molecules were placed on Parafilm substrate; the experimental setup is sketched in FIG.~\ref{fig:1}(A). The initial concentration $C$ of the HFBI molecules is kept in the range $\SI{2}{\micro\mole\per\liter}\leq C\leq\SI{4}{\micro\mole\per\liter}$.
The droplets are monitored using an optical tensiometer and their profile captured using a CCD camera; experimental images are processed using ImageJ and the library openCV \cite{Note1}.
Sessile droplets are found to develop a flattened region during evaporation 
that is made by a crumpled HFBI monolayer, as shown in FIG.~\ref{fig:1}(C, D). Pendant droplets are obtained by quickly inverting the substrate after droplet deposition; during evaporation they display a circumferential wrinkling in the vicinity of the contact line, see FIG.~\ref{fig:1}(C, D).

The crumpling of the flattened region indicates that the elastic film is not flat in its relaxed state and exhibits a non-zero Gaussian curvature. (Similar crumpling has been observed previously in polydopamine-stabilized droplets deflated by syringe suction \cite{Abe_2017}, in  polypyrrole droplets\cite{Kim_2019}, as well as in water drops partially covered by a flat polystyrene sheet \cite{King2012}.)
Moreover, the formation of wrinkles in pendant droplets suggests that the HFBI film is also present on the entire free surface of the droplets.
Therefore, a key first question is to identify the critical threshold for  the onset of encapsulation of HFBI droplets.

\begin{figure}
\includegraphics[width=0.9\columnwidth]{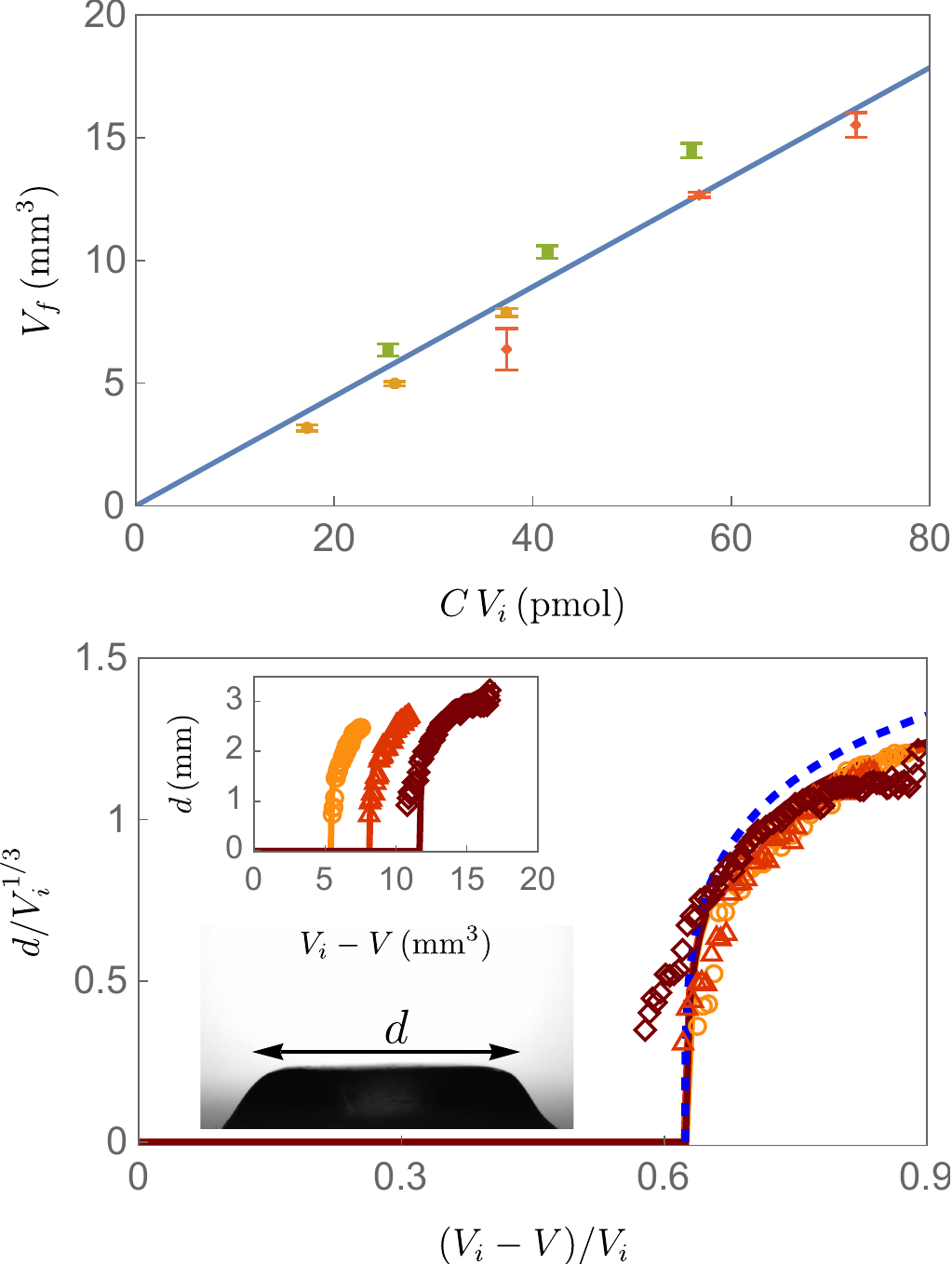}
\caption{(Top): Experimental data for the droplet volume $V_f$ at which the sessile droplets start to flatten as a function of the total number of HFBI molecules contained in the droplet (the product of initial concentration $C$  and  initial droplet volume $V_i$). The orange, green and red dots represent the data for $C = 2,\,3,\,4\,\SI{}{\micro\mole\per\liter}$, respectively. The blue solid line shows the linear fit of the experimental data. (Bottom): The diameter, $d$, of the flat spot increases with the volume of liquid lost to evaporation, $V_i-V$. Raw measurements for droplets with $C = \SI{2}{\micro\mole\per\liter}$ and $V_i = 8.63,\,13.05,\,18.68\,\SI{}{\milli\meter^3}$ (yellow, red, and brown, respectively) shown in the inset collapse onto a master curve when lengths are rescaled by $V_i^{1/3}$. The blue dashed curve shows the prediction of a purely geometrical model, see Eq.~(S22) of \cite{Note1}.}
\label{fig:2}
\end{figure}

Analysing the experimental images, we find that the critical
volume $V_f$ at which the flattening of sessile droplets is first observed is linearly proportional to the initial amount of HFBI contained in the droplet $CV_i$ (with $C$ the initial concentration and $V_i$ initial volume), see FIG.~\ref{fig:2}, top. Assuming that flattening starts soon after the encapsulation of droplets, the plot of FIG.~\ref{fig:2} suggests that the encapsulation takes place when the bulk concentration of HFBI reaches a given threshold, such that the interfacial concentration (assumed in equilibrium with the bulk) reaches a critical value. Thus, we
adopt the following phenomenological law for computing the volume $V_e$ at
which the droplet is encapsulated:
\begin{equation}
\label{eq:Ce}
V_e = \frac{C V_i}{C_e}.
\end{equation}
where $C_e$ is a constant representing the critical bulk concentration at which the encapsulating membrane is formed.
Moreover, in FIG.~\ref{fig:2} (bottom) we show the diameter $d$ of the flattened region in sessile droplets as a function of the evaporated volume. Interestingly, the experimental  curves display a universal scaling law, since all dimensionless data collapse onto a master curve taking a length scale $V_i^{1/3}$, suggesting a universal geometrical scaling independent of gravity \cite{Note1}.

We now introduce a theoretical model to unravel the key mechanisms of the observed droplet morphology before and after the formation the elastic film. We distinguish two main phases during the evaporation process.
Before the formation of the elastic film, the shape of the
drop is dominated by the interplay between gravity and capillarity. 
We assume axisymmetry and we fix the center of a Cartesian frame $(x,\,y,\,z)$
at the apex of the drop. Let $(r(t,\,S),\,z(t,\,S))$ be the curve describing the
shape of the droplet at time $t$, where $r$ is the radial distance from the $z$
axis and $S$ is the arc-length of the curve.

In our experiments we use  droplets of characteristic length  $V_i^{1/3}$ on the order of millimeters. The Bond number of the droplet (whose sign is positive or negative for sessile or pendant droplets, respectively) $\mathrm{Bo} = \pm\rho g V_i^{2/3}/\gamma$, is therefore $O(1)$ in all the experiments. Hence, gravitational forces are of the same order as capillary forces. The adsorption of HFBI molecules at the interface leads them to self-assemble to form a highly-ordered protein film, with a stretching modulus of similar order to the surface tension \cite{Knoche_2013,al2017study}.

Because of gravity, the pressure inside the drop is $p = p_T - \rho g z$ where $p_T$ is the pressure at
$z=0$. Thus, before the elastic membrane forms, the droplet shape is governed by the Young--Laplace equation at the interface 
\begin{equation}
\label{eq:YL}
\rho g z - p_T= \gamma \kappa.
\end{equation}
where $\gamma$ is the surface tension and $\kappa$ is twice the mean curvature. If $\phi$ denotes the angle between the tangent and the radial direction, $\kappa = \pm \left(d\phi/dS + \sin\phi/r\right)$, where the sign is positive or negative for sessile or a pendant droplets, respectively.
Using as initial conditions $r(t,\,0)=z(t,\,0)=0$, we numerically solve the Young--Laplace equation~\eqref{eq:YL} for a fixed value of $p_T$ in the interval $S\in(0,\,S_{end})$ with $r(t,\,S_{end})$ equal to the fixed contact radius $r_c$ \cite{Note1}.

Once the bulk HFBI concentration reaches the critical value $C_e$, i.e.~the droplet reaches the critical volume $V_e$ given by equation \eqref{eq:Ce}, HFBI molecules at the free surface  self-organize into a monolayer with a hexagonal structure, as shown in FIG.~\ref{fig:1} (B). In this second phase, the shape of the droplet is dictated by the interplay between the film elasticity  and the gravitational forces on the droplet. Since the bending modulus $B$ of the HFBI film is proportional to the cube of the membrane thickness (proportional to the size of a HFBI droplet), the bendability $\gamma V_i^{2/3}/B$ is much larger than $1$ and  the elastic film can be modeled as an elastic membrane \cite{Davidovitch_2011}.

Denoting by $r_0$ and $z_0$ the radial and axial
coordinates, respectively, of the membrane cross-section as  the HFBI film forms, let $s_0\in[0,\,L]$ be the arc-length of this initial curve. Its current configuration during evaporation, is described by the current  coordinates $r$ and $z$, with the interface parametrized by the arc-length  $s=s(s_0)$. 

Taking account of both elastic and capillary
forces, the membrane assumption for the meridional and hoop stresses $\tau_s$ and $\tau_\theta$  gives
\begin{equation}
\label{eq:tensions}
\left\{
\begin{aligned}
&\tau_s = \frac{E}{1-\nu^2} \frac{1}{\lambda_\theta}\left[(\lambda_s-1)+\nu(\lambda_\theta-1)\right]+\gamma,\\
&\tau_\theta = \frac{E}{1-\nu^2} \frac{1}{\lambda_s}\left[(\lambda_\theta-1)+\nu(\lambda_s-1)\right]+\gamma,\\
\end{aligned}
\right.
\end{equation}
where $E$ is the 2D Young's modulus, $\nu$ is the 2D Poisson's ratio, while
$\lambda_s=\upd s/\upd s_0$ and $\lambda_\theta = r/r_0$ are the meridional and hoop
stretches, respectively \cite{Libai_1998}. We remark that the elastic membrane is in tension at the
instant of film formation (when $\lambda_s = \lambda_\theta = 1$) thanks to capillary forces.

Assuming quasi-static deformations,  the following force balance equations must be imposed:
\begin{equation}
  \label{eq:balance}
  \left\{
  \begin{aligned}
  &\frac{\upd}{\upd s}(r\tau_s)-\tau_\theta\cos\phi=0,\\
  &\kappa_s\tau_s+\kappa_\theta\tau_\theta=\rho g z - p_T
  \end{aligned}
  \right.
\end{equation}
where $\kappa_s$ and $\kappa_\theta$ are the principal curvatures
along the meridional and hoop directions, respectively.
In \eqref{eq:balance}, the first equation represents in-plane equilibrium, while the latter represents equilibrium normal to the interface and hence is the elastic counterpart of the Young--Laplace
equation~\eqref{eq:YL}. Setting the boundary conditions $r(0)=\phi(0)=0$, the ordinary differential system is numerically solved by fixing the value of $\tau_s(0)$ and by
performing a shooting method with the parameter $p_T$ until the boundary
condition $r(t,\,L)=r_c$ is satisfied \cite{Note1}.

\begin{figure*}[ht!]
\includegraphics[width=0.9\textwidth]{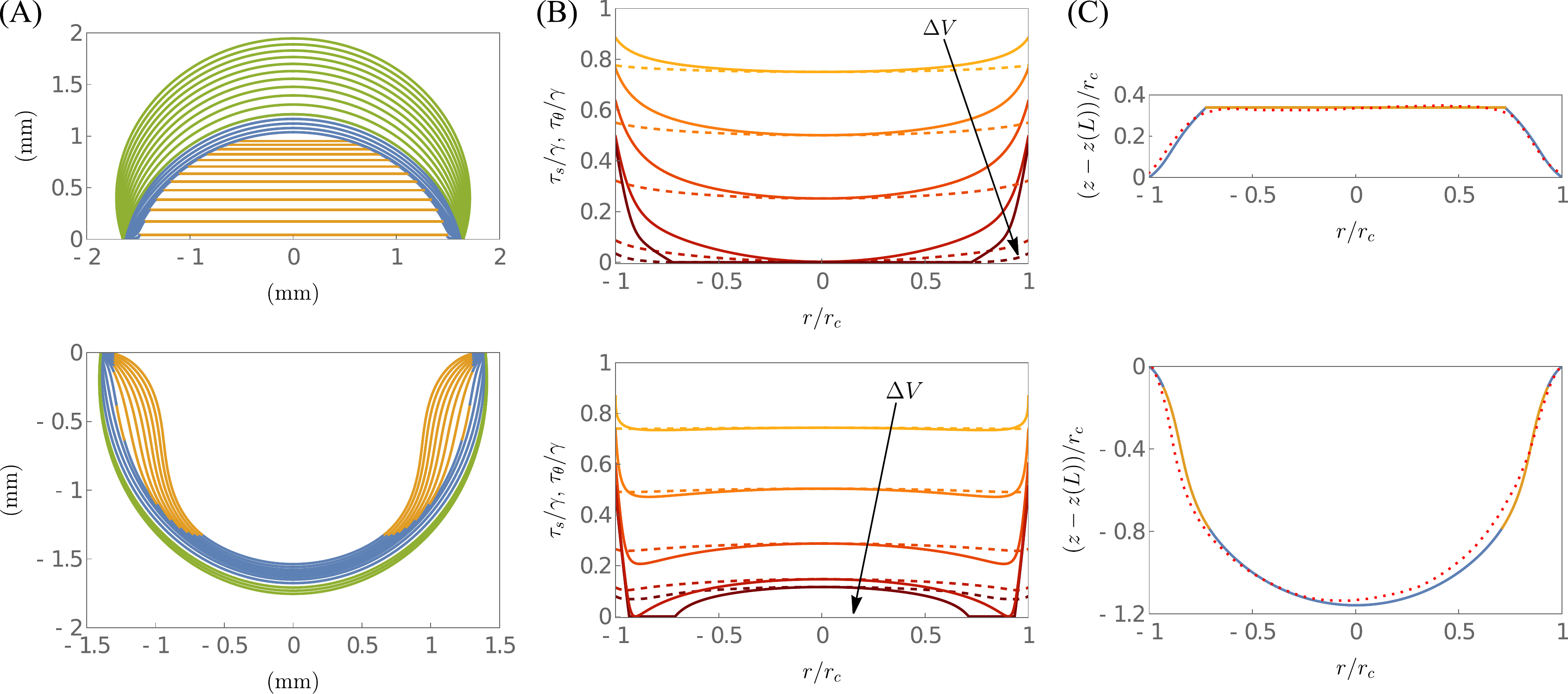}
\caption{(A): Theoretical predictions of the droplet profiles during evaporation of a sessile (top, $V_i = \SI{13.05}{\milli\meter^3}$, $C = \SI{2}{ \micro\mole\per\liter}$, $\mathrm{Bo}=0.99$) and a pendant droplet (bottom, $V_i = \SI{7.77}{\milli\meter^3}$, $C = \SI{4}{\micro\mole\per\liter}$, $\mathrm{Bo}=-0.70$). The green solid  curves  are obtained from the fluid equation \eqref{eq:YL}, which remains valid while the bulk concentration $C<C_e$; while for $C\geq C_e$ the elastic  equations \eqref{eq:tensions}, \eqref{eq:balance} are solved. Blue and orange solid curves indicate the  elastic regions and wrinkled/crumpled regions of the interface, respectively. (B):  Dimensionless meridional (dashed curves) and hoop stress (solid curves)
versus the dimensionless radial position $r/r_c$ at different volumes $V$ ($V=3.57,\,4.91,\,5.26,\,5.60,\,\SI{5.93}{\milli\meter^3}$ for the sessile droplet, while $V=5.77,\,6.15,\,6.43,\,6.76,\,\SI{7.10}{\milli\meter^3}$ for the pendant droplet), the arrows indicate the direction of increasing  evaporated volume $\Delta V$. (C): Predicted (solid curves) and experimental droplet profiles (dotted curves) at a given  time for a sessile (top, $V = \SI{3.57}{\milli\meter^3}$) and a pendant droplet (bottom, $V = \SI{5.77}{\milli\meter^3}$). Here color shows the predicted wrinkled/crumpled (yellow) or tensile (blue) regions.}
\label{fig:3}
\end{figure*}

As water evaporates, surface tension is counteracted by compressive elastic stresses within the membrane, depending on the dimensionless elastocapillary ratio $E/\gamma$. Since the film has negligible bending stiffness, it buckles immediately upon compression forming wrinkles (if the compression is uniaxial) or crumples (if the compression is biaxial). To understand the shape evolution beyond the onset of crumpling/wrinkling,  we apply a far-from-threshold analysis  \cite{Davidovitch_2011,Knoche_2013}, assuming that the membrane undulations have  small amplitude and short wavelength.  In particular, we describe the elastic
film through an axisymmetric pseudosurface $(r(t,\,s_0),\,z(t,\,s_0))$ around
 the buckled (non-axisymmetric) mid-surface.

For sessile droplets, our theory predicts that the apex of the membrane becomes crumpled during evaporation. We assume that the stress is completely released therein, i.e.~$\tau_s = \tau_\theta = 0$ in a portion of the membrane $s_0\in[0,\,s_f]$; since both $\tau_s$ and $\tau_\theta$ vanish in this region, the HFBI film cannot  sustain a pressure difference across it, and so we must have $p_T=0$ and the interface is flat. Hence, a flat spot appears at the apex of the droplet, as observed experimentally. Beyond the flat spot, \eqref{eq:balance} continues to hold. 
It is possible to show that the actual radius of the flattened
region is given by $r(t,\,s_f) = E/(E+\gamma-\gamma\nu)r_0(t,\,s_f)$  \cite{Note1}. Using this
boundary condition, we numerically integrate the system \eqref{eq:balance} for
$s_0\in[s_f,\,L]$ by fixing $s_f$ and using a shooting method to find the
value of $\phi(t,\,s_f)$ at which the boundary condition $r(t,\,L) = r_c$ is
satisfied. The resulting diameter of the crumpled region collapses on a master curve when plotted versus the evaporated volume,  in excellent agreement with the experimental findings (see FIG~\ref{fig:2}, bottom). This collapse can be explained using a geometrical argument that explicits the closeness of the droplet shape to a spherical cap \cite{Note1}. A similar numerical scheme is applied to model pendant droplets \cite{Knoche_2013}; here we find that only the hoop stress vanishes in a region $[s_a,\,s_b]\subseteq[0,L]$; radial wrinkles therefore form and the balance equations \eqref{eq:balance} are modified
 by setting $\tau_\theta=0$ for $s_0\in[s_a,\,s_b]$ \cite{Note1}.

In the numerical simulations, we set $E=\SI{400}{\milli\newton\per\meter}$, $\nu=0.5$, $\gamma=\SI{55}{\milli\newton\per\meter}$, and $C_e=\SI{4.17}{\micro\mole\per\liter}$. 
The numerical results display excellent  agreement with  experimentally-determined profiles for both sessile and pendant droplets. Two illustrative examples of the morphological evolution of the pendant and the sessile droplets during evaporation are shown in FIG.~\ref{fig:3}. We remark that, despite the morphological differences between the pendant and the sessile droplets, in both the cases the hoop stress exhibits huge variations along the radial direction and is maximal near the contact line (namely for $s_0$ close to $L$), while the meridional stress is nearly constant. Specifically,  at the apex the stress is a locally convex or concave function of the radial coordinate if the droplet is sessile or pendant, respectively, see FIG.~\ref{fig:3}. This stress distribution is key to whether crumples or wrinkles form: for both sessile and pendant drops, the stress is isotropic close to the apex, i.e.~$\tau_s\approx\tau_\theta$. For sessile droplets this isotropic stress is a  global minimum so that both stresses become compressive simultaneously, leading to crumpling at the apex. For pendant droplets  both stresses are locally maximal at the apex, with a global minimum in $\tau_\theta$ close to the contact line --- compression occurs with $\tau_\theta=0$, $\tau_s>0$ and  wrinkles appear.

\begin{figure}
\includegraphics[width=0.85\columnwidth]{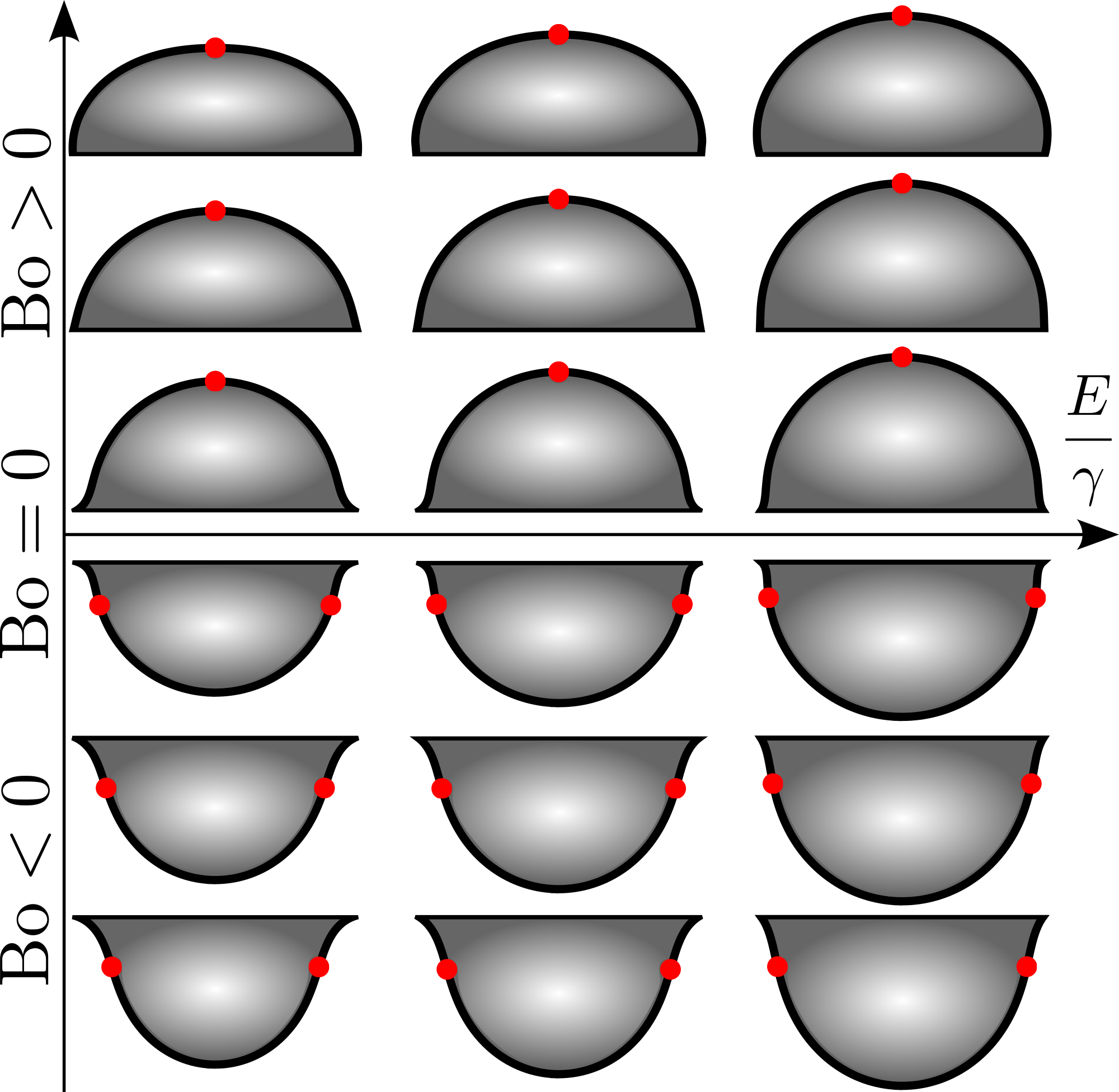}
\caption{Morphological diagram of the droplets with respect to the dimensionless parameters $\mathrm{Bo}$ and $E/\gamma$ at the onset of wrinkling ($\mathrm{Bo}<0$) and crumpling ($\mathrm{Bo}>0$). Red dots indicate the nucleation points of wrinkling/crumpling. We set $\mathrm{Bo}=\pm 0.07,\,\pm 0.7,\,\pm1.4$, with constant initial volume, and $E/\gamma = 2.7,\,3.6,\,7.3$.}
\label{fig:4}
\end{figure}

Figure~\ref{fig:4} shows a morphological diagram for evaporating HFBI-water droplets in terms of the dimensionless parameters $\mathrm{Bo}$ and $E/\gamma$. This plot emphasizes the unusual change in behavior seen as $\mathrm{Bo}$ changes sign: the two drop shapes with $\mathrm{Bo}=\pm0.07$ and fixed $E/\gamma$ are very similar (as should be expected for small $|\mathrm{Bo}|$). Nevertheless, the location at which wrinkles/crumples are nucleated is very different in the two cases. FIG.~\ref{fig:4} also shows that at fixed drop volume (fixed $\mathrm{Bo}$) a smaller volume decrease is required to induce compression as the dimensionless elasticity, $E/\gamma$, increases.

In this Letter we have presented a theoretical framework to predict the morphological changes observed  during the evaporation of  sessile and pendant HFBI-water  droplets. Our experiments show that the HFBI molecules self-assemble creating a encapsulating monolayer when the bulk concentration of HFBI molecules reaches a critical value, $C_e$. Upon further evaporation, the stress in this layer becomes compressive and the layer wrinkles or crumples. Surprisingly, these changes strongly depend on the direction of gravity, with a discontinuous transition between a flat spot at the apex, and wrinkles near the contact line as $\mathrm{Bo}$ changes sign. This distinction is observed experimentally and quantitatively explained by our gravito-elasto-capillary model. The droplet morphology results from the interplay between elasticity, capillarity, evaporation, and gravity, as summarized in  FIG.~\ref{fig:4}. This new understanding may be used to propose effective mechanisms to encapsulate droplets  that change their shape on demand.  Controlling the morphological changes that take place as the volume of the encapsulated droplet decreases (through either evaporation or the controlled release or removal of the encapsulate \cite{Ma_2014}) has many applications. For example, change of shape  can be used to target adhesion on surfaces \cite{Decuzzi_2008}, to regulate drag in fluid-structure interaction \cite{Terwagne_2014}, or to control depinning of heated droplets in microgravity conditions~\cite{kumar2020sessile}. Future efforts will be devoted to the study of picoliter HFBI droplets, as well as droplets resting on a nonrigid substrate (e.g.~elastic or liquid substrates).

\begin{acknowledgments} This work has been partially supported by MUR (PRIN grant MATH4I4 2020F3NCPX), by Regione
Lombardia (NEWMED  grant,  ID:  117599, POR FESR 2014-2020), by GNFM -- INdAM, by the Academy of Finland (Center of Excellence Program (2022-2029) in Life-Inspired Hybrid Materials (LIBER) project number 346109). We also acknowledge the provision of facilities and technical support by Aalto University at the OtaNano Nanomicroscopy Center. \end{acknowledgments}

%

\end{document}



\title{Supplementary material for ``\emph{Flattened and wrinkled encapsulated droplets: Shape-morphing induced by gravity and evaporation}"}

\author{Davide Riccobelli}
\affiliation{MOX -- Dipartimento di Matematica, Politecnico di Milano, Italy}
\author{Hedar H. Al-Terke}
\affiliation{Department of Applied Physics, Aalto University School of Science, Finland}
\affiliation{Center of Excellence in Life--Inspired Hybrid Materials (LIBER), Aalto University, Finland}
\author{P\"{a}ivi Laaksonen}
\affiliation{HAMK Tech,  Häme University of Applied Sciences}
\author{Pierangelo Metrangolo}
\affiliation{Department of Chemistry, Materials, and Chemical Engineering “Giulio Natta”, Politecnico di Milano, Italy}
\affiliation{Department of Applied Physics, Aalto University School of Science, Finland}
\affiliation{Center of Excellence in Life--Inspired Hybrid Materials (LIBER), Aalto University, Finland}
\author{Arja Paananen}
\affiliation{VTT Technical Research Centre of Finland Ltd, Finland}
\author{Robin H. A. Ras}
\affiliation{Department of Applied Physics, Aalto University School of Science, Finland}
\affiliation{Center of Excellence in Life--Inspired Hybrid Materials (LIBER), Aalto University, Finland}
\author{Pasquale Ciarletta}
\affiliation{MOX -- Dipartimento di Matematica, Politecnico di Milano, Italy}
\author{Dominic Vella}
\affiliation{Mathematical Institute, University of Oxford, Woodstock Rd, Oxford, OX2 6GG, UK.}

\date{\today}

\maketitle

This Supplementary material contains further details of the experimental procedures used in this work (\S \ref{sec:Methods}) as well as a dimensional analysis of the relative importance of the various physical effects  at work in the experimental problem (\S\ref{sec:DimAnalysis}) and the details of the theoretical model used to predict the interface shape (\S\ref{sec:Theory}).

\section{Materials and methods \label{sec:Methods}}

\subsection{Droplet preparation}
HFBI Hydrophobin protein is produced and purified as described by Linder \emph{et al.}~\cite{Linder_2001} and dispersed in Milli-Q water, which is water purified using a Milli-Q system (Millipore) with resistivity around \SI{18.2}{\mega\ohm\centi\meter}.

 A water droplet containing HFBI hydrophobin molecules is placed on a parafilm substrate using a micro-pipette. The droplet is monitored using an optical tensiometer (Biolin Attention Theta Goniometer, Finland) and the profile of the droplet captured using a CCD camera at a frame rate of $0.05\mathrm{~fps}$. All the experiments are performed at room temperature (around $\SI{20}{\celsius}$) and with a relative humidity of $25\%$.
In the experiments with pendant droplets, a water droplet containing HFBI molecules is placed on the parafilm substrate in the sessile state and the film is quickly inverted to yield the pendant state  (see FIG.~1 of the main text).

\subsection{Image processing}
\begin{figure}
    \centering
    \includegraphics[height=0.3\textwidth]{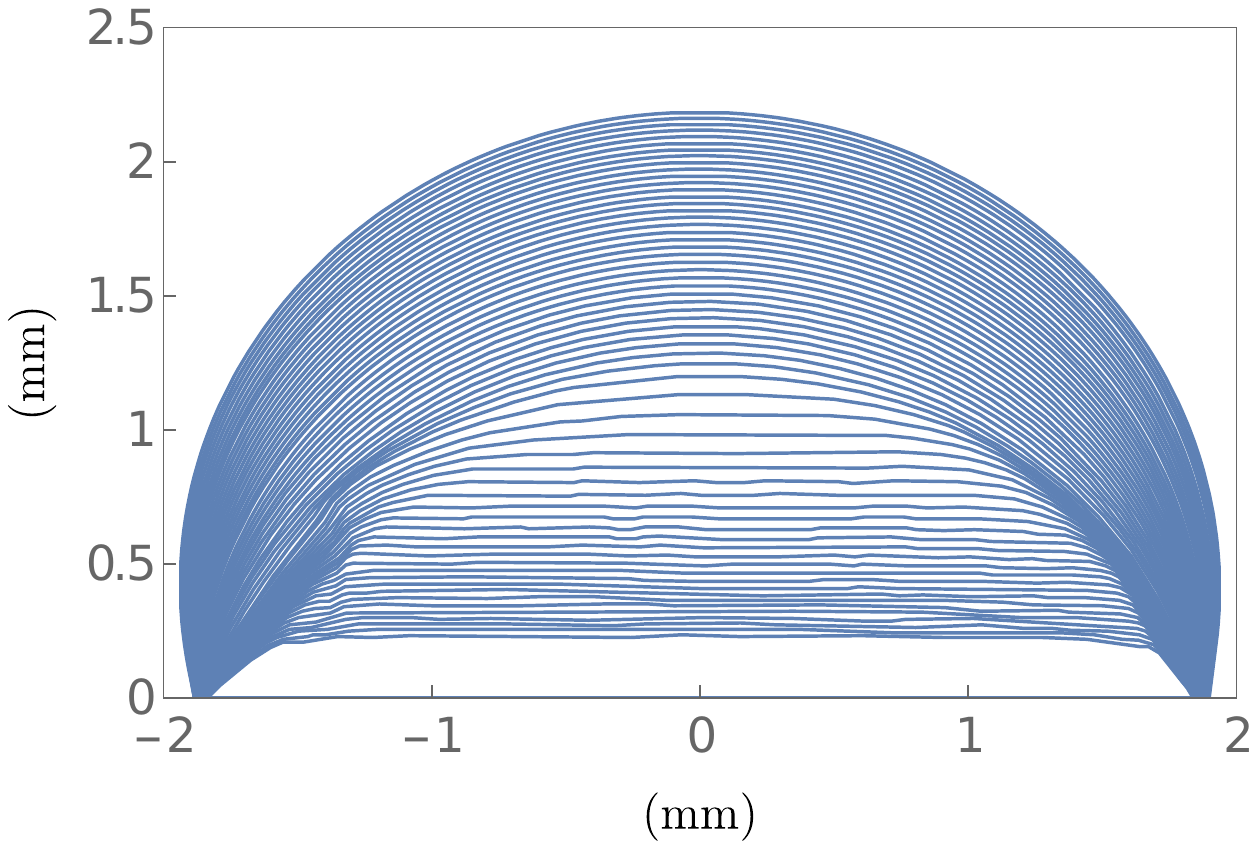}\quad
    \includegraphics[height=0.3\textwidth]{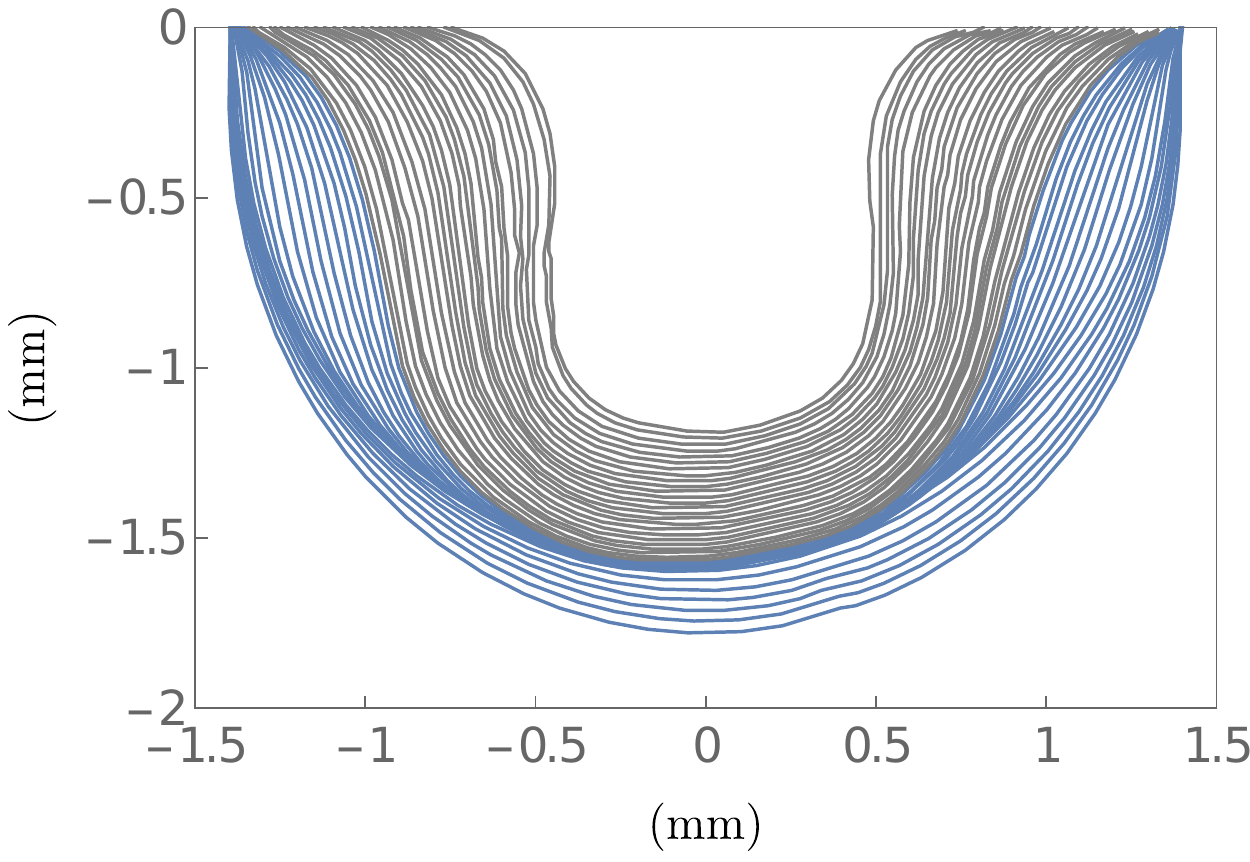}
    \caption{Plots of the extracted experimental profiles for (left) a sessile droplet with initial volume $V_i = \SI{18.68}{\milli\meter^3}$ and initial concentration $C = \SI{2}{\micro\mole\per\liter}$ and (right) a pendant droplet with initial volume $V_i = \SI{7.77}{\milli\meter^3}$ and initial concentration $C = \SI{4}{\micro\mole\per\liter}$. Individual curves show the shape extracted at intervals of one minute from the start of evaporation.  We note that the contact line is initially pinned in pendant droplets, but ultimately depins after some level of evaporation. We identified contact line depinning as when the contact radius has diminished by more than $4\%$ from its initial value; we distinguish profiles in which the contact line is depinned by using gray curves (rather than blue curves, which represent the profiles of pinned droplets).}
    \label{fig:exp_profiles}
\end{figure}

Experimental images are processed with several software packages to extract the profile of the evaporating drops. First, the pictures are converted to binary images using \textsc{ImageJ} and processed using the library \textsc{MorphoLibJ} \cite{Legland_2016}. The droplet interface is determined by extracting the boundaries using the Python library \textsc{OpenCV}. Finally, the drop profiles obtained in this way are processed using \textsc{Mathematica} to extract information such as the volume of the drop, the diameter of the flat spot of the surface, and the height of the drop. FIG.~\ref{fig:exp_profiles} shows examples of the shape evolution determined in this way.

\begin{figure}[t!]
    \centering
    \includegraphics[width=0.5\textwidth]{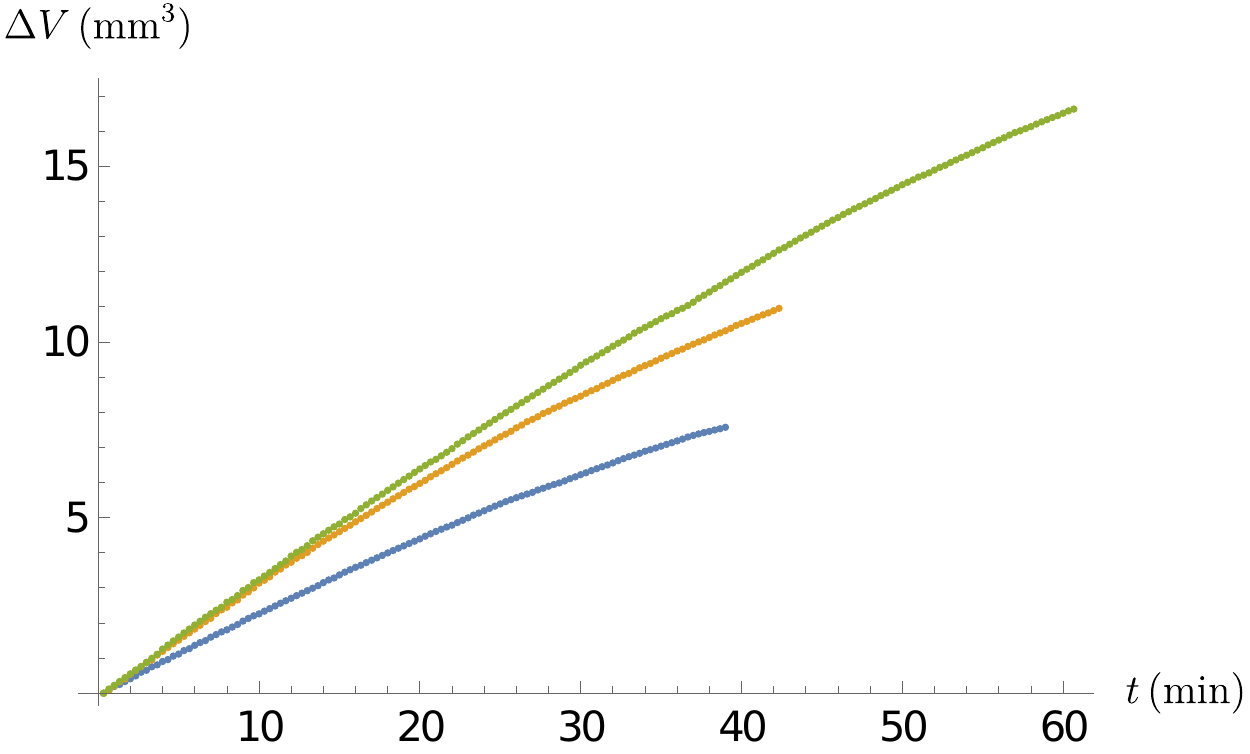}%
    \includegraphics[width=0.5\textwidth]{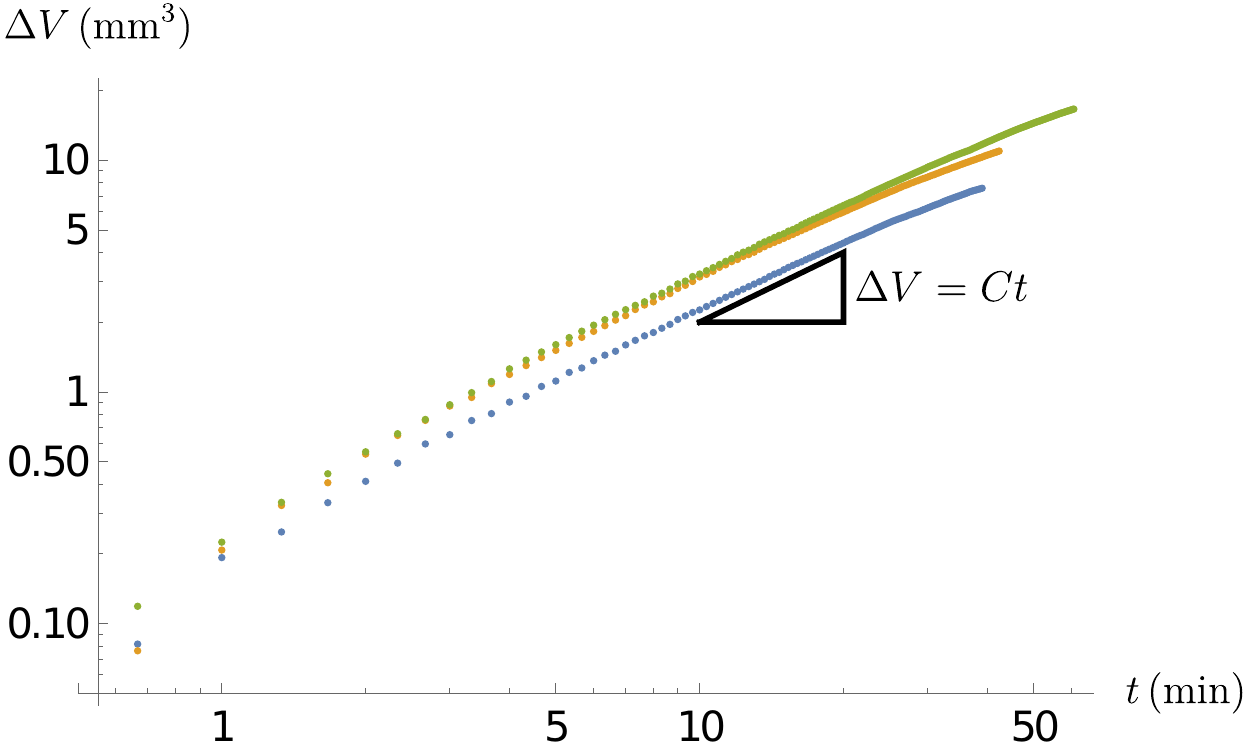}
    \caption{(Left) Evaporated volume, $\Delta V$, plotted as function of time for three droplets with initial volume $V=8.63,\,13.05,\,\SI{18.68}{\milli\meter\cubed}$ (blue, orange, and green dots, respectively) and initial concentration $C=\SI{2}{\micro\mole\per\liter}$. (Right) The same data plotted using logarithmic axes. The triangle indicates the slope expected for a linear relation (i.e.~a constant evaporation rate).}
    \label{fig:ev_vol}
\end{figure}

Time lapse videos of the evolution of droplet shape through evaporation are shown for the sessile and pendant configurations in the movies \texttt{SM\_video\_sessile.avi} and \texttt{SM\_video\_pendant.avi}, respectively. In these movies, images are presented at intervals of $20\mathrm{~s}$.
We observe that the evaporation rate of the droplets is approximately constant in time and is not significantly influenced by the formation of the HFBI membrane, as shown in FIG.~\ref{fig:ev_vol}.

\subsection{Sample preparation for Atomic Force Microscopy (AFM)}
A droplet of $\SI{100}{\micro\liter}$ of HFBI solution ($\SI{0.2}{\gram\per\liter}$) was placed on a parafilm substrate for about one hour to get a flattened area on the top of the droplet and to test the assembly of a HFBI monolayer at the interface. After that, a highly ordered pyrolytic graphite (HOPG) substrate was brought into contact with the top of the droplet. By gently washing it with $\SI{200}{\milli\liter}$ of Milli-Q water the excess HFBI molecules were taken out. The sample was then placed on the AFM stage and, after waiting around $\SI{30}{\minute}$ to ensure the sample was dry, the sample was imaged. The full experimental pipeline is sketched in FIG.~\ref{fig:AFM}).

\subsection{Atomic force microscopy}
\begin{figure}[!h]
    \centering
    \includegraphics[width=0.7\textwidth]{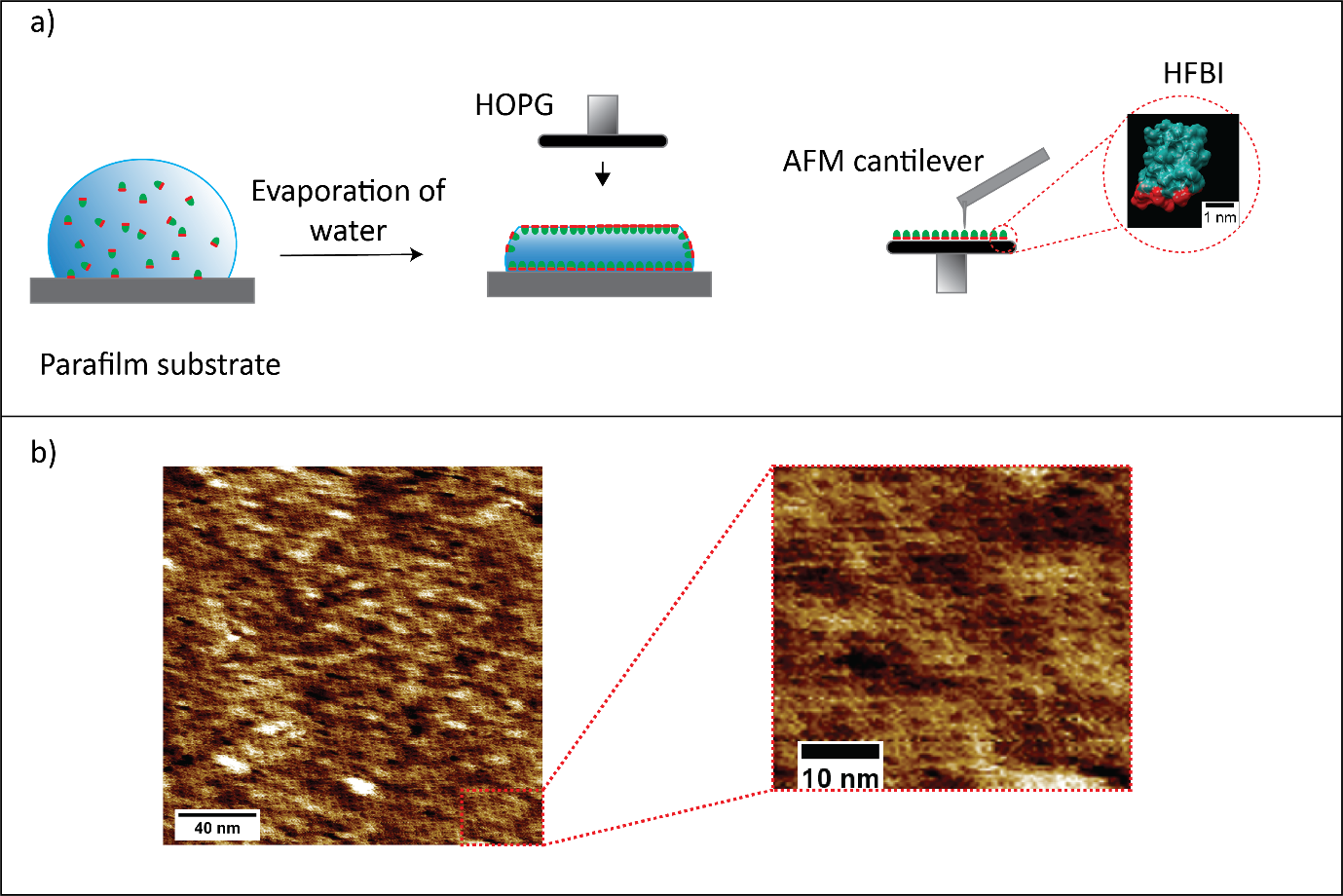}
    \caption{a) Schematic showing the preparation of the sample for AFM imaging. b) AFM image of a hydrophobin monolayer formed at an air--water interface after being transferred to an HOPG surface. }
    \label{fig:AFM}
\end{figure}

A Dimension Icon AFM (Bruker AXS, France; formerly Veeco) was used to carry out the atomic force microscopy (AFM) experiments. The device is equipped with a ScanAsyst-air cantilever (sharp silicon nitride tip with a nominal radius of $\SI{2}{\nano\meter}$ for PeakForce Tapping in air). With a resolution of $256\,\mathrm{pix}\times256\,\mathrm{pix}$ and scan rate of $\SI{1}{\kilo\hertz}$, the AFM scanning was performed for a $\SI{100}{\nano\meter}\times\SI{100}{\nano\meter}$ scan area. ScanAsyst Auto control was set to ``individual'' for the sample with PeakForce Amplitude of $\SI{170}{\nano\meter}$. The spring constant and peak force frequency were $\SI{0.4}{\newton\per\meter}$ and $\SI{2}{\kilo\hertz}$ for all samples. Individual scans for each sample were taken at multiple locations on the surface. An example scan is shown in FIG.~\ref{fig:AFM}b.

\section{Dimensional analysis\label{sec:DimAnalysis}}

\subsection{Dimensional parameters}
The fundamental  physical parameters of this system, used to develop our model, are collected in  Table \ref{table1}. Other fundamental parameters can be computed as derived quantities for the physics at hand. In particular, the apparent diffusivity $D_{H}$ of HFBI in water is given by the Stokes--Einstein relation \cite{krivosheeva2013kinetic}:
$$D_{H}= \frac{\kappa_B T_0}{6 \pi \mu  a }= \SI{1.4e-10}{\meter^2\per\second}, $$ 
where $\kappa_B= \SI{1.38e-23}{\square\meter\kilo\gram\per\square\second\per\kelvin}$ is the Boltzmann constant. 
At room temperature, $T_0\approx 293\mathrm{~K}$, the saturated water vapor pressure $p_v$ can be computed as \cite{patience2017experimental}:
$$ p_v= \SI{610.78}{\pascal} \ {\rm exp}\left( \frac{17.26(T_0-\SI{273.16}{\kelvin})}{ T_0 -\SI{34.86}{\kelvin} }  \right)= \SI{2325.98}{\pascal},  $$
allowing us to compute the vapour saturation concentration $c_v$ as:
$$c_v= \SI[per-mode=fraction]{0.002166}{\kelvin\square\second\per\square\meter} \frac{p_v}{T_0}= \SI{1.72e-2}{\kilo\gram\per\meter\cubed}. $$
\begin{table}[!t]
\begin{tabular}{|l|c|c|c|}
\hline
\textbf{Parameter} &  \textbf{Symbol}& \textbf{Value} & \textbf{REF}. \\ \hline
water density & $\rho$ & \SI{1e3}{\kilo\gram\per\meter\cubed} & \cite{korson1969viscosity} \\ \hline
HFBI density & $\rho_{H}$ & \SI{1.13e3}{\kilo\gram\per\meter\cubed} & \cite{Yamasaki_2016} \\ \hline
humidity & $h$ & $25\%$ &Measured \\ \hline
surface tension droplet-air & $\gamma$ & \SI{55e-3}{\newton\per\meter} & \cite{al2017study} (FIG.~22 p.34)\\ \hline
size of HFBI particle & $a$ & \SI{1.5}{\nano\meter}  & \cite{paananen2003structural} \\ \hline
water viscosity & $\mu$ & \SI{e-3}{\pascal \second}  & \cite{korson1969viscosity} \\ \hline
thermal conductivity water & $k_w$ & \SI{0.58}{\watt\per\meter\per\kelvin} & \cite{nieto1986standard}  \\ \hline
initial room temperature & $T_0$ & \SI{293}{\kelvin} & Measured  \\ \hline
diffusivity of vapor in air  & $D_w$ & \SI{26.1e-5}{\meter\squared\per\second} & \cite{welty2009fundamentals}  \\ \hline
latent heat of vaporization  & $H_v$ & \SI{2260}{\kilo\joule\per\kilo\gram} & \cite{perry1950chemical}  \\ \hline
coefficient of variation of surface tension
& $\beta=d\gamma/d T$& 
\SI{-0.16}{m\newton\per\kelvin\per\meter}  & \cite{Gittens1969}  \\ \hline
\end{tabular}
\caption{Physical parameters of the system that are used to develop the model presented here.}
\label{table1}
\end{table}

\subsection{Characteristic parameters}

We start by introducing the main characteristic quantities that enter  the  model of the system  presented in the main paper. We are particularly focused on understanding the forces that are involved in determining the shape of the droplet, since this is the focus of the paper.

\subsection{Gravitational effect on individual molecules}
The previous suggestion has been that the effect of gravity on individual molecules is important. Here we estimate the importance of this effect by estimating the sedimentation length, $\ell_s$, that is the height over which the gravitational potential energy of the molecule would become comparable to the thermal energy, $\kappa_BT$, \cite{Piazza2012}. For systems that are small compared to $\ell_s$, thermal energy is expected to overwhelm the effects of gravity.

The sedimentation length is calculated as
\[
    \ell_s=\frac{\kappa_BT_0}{\Delta\rho g a^3}
\]
where $\Delta\rho=\rho_H-\rho=\SI{130}{\kilo\gram\per\cubic\meter}$ is the density difference between HFBI and water, $g$ is the acceleration due to gravity and $a$ is the molecular size. We therefore have that $\ell_s\approx\SI{1}{\kilo\meter}$. As this kilometre length scale is at least five orders of magnitude larger than any relevant length scale in the problem, it is safe to neglect the effect of gravity on individual molecules. Instead, we turn to consider the effect of gravity on the droplet itself.

\subsubsection{Gravity versus surface tension for the droplet}

The shape of a stationary liquid droplet, in equilibrium, is controlled by a balance between surface tension and gravity. This balance introduces a characteristic length scale,  the capillary length  $\ell_c$, which is defined as:
$$ \ell_c= \sqrt{\frac{\gamma}{\rho g}}= \SI{2.37}{\milli\meter},$$ 
where $g$ is the acceleration due to gravity. We note that $\ell_c$ is of the same order as the contact radius $r_c$  of the droplet and, further, that the Bond number
$$\mathrm{Bo}=\pm\left(\frac{V_i}{ \ell_c^3}\right)^{2/3} \approx \pm 1, $$
where the plus or minus sign depends on the orientation of the droplet with respect to gravity. Specifically, we take a positive sign for sessile droplets, while $\mathrm{Bo}$ is negative in pendant droplets.
Since $|\mathrm{Bo}|\approx 1$, gravity cannot be neglected when computing the shape of the droplet --- the effects of the droplet's weight and surface tension are expected to be of similar order.

\subsubsection{Fluid flow}

Having seen that the effects of gravity and capillarity are of the same order of magnitude within the droplet, the question then arises of whether the shape of the droplet is determined by these forces (together with elasticity), or whether the fluid flow induced by evaporation plays an important role.

We begin by estimating the characteristic velocity, $v_c$, of the capillary flow driven by evaporation: the evaporative flux is driven by steady-state diffusion \cite{Deegan_1997,bhardwaj2009pattern}, and hence has magnitude $j\sim D_w c_v/\ell_c$
 where we assume that diffusion occurs over the typical size of the drop (itself comparable to the capillary length) and the typical vapour concentration is $c_v$. Since this represents a mass flux per unit area (recall that $c_v$ is a mass per volume) it yields a capillary velocity:
 $$v_c=\frac{D_w  c_v }{\rho \ell_c}\approx \SI{1.89}{\micro\meter\per\second}.$$
The Reynolds number associated with this flow is
 $$ \mathrm{Re}= \frac{\rho v_c \ell_c}{\mu}\approx\SI{4e-3}{},$$
 so that inertia is negligible in comparison to viscous forces, while the relative importance of viscous to capillary forces is controlled by the capillary number
 $$\mathrm{Ca}= \frac{\mu v_c}{\gamma}= \SI{3.4e-8}{}.$$ This shows that viscous forces are negligible in comparison to capillary forces, and the droplet shape is therefore governed by static considerations, as assumed in the model of the main paper.
 
 \subsubsection{Temperature change and Marangoni effects}
 The evaporation of liquid with the flux $j$, may also induce a temperature change --- the rate at which heat energy is lost through evaporation is $j H_v $ with $H_v$ (the latent heat of vaporization), heat that must be supplied by thermal conduction through the droplet driven by a temperature change $\Delta T\sim jH_v/(k_w/\ell_c)$, so that the characteristic temperature $\Delta T$ is given by
$$\Delta T=\frac{D_w H_v c_v}{k_w}= \SI{17.51}{\kelvin}. $$

The most important consequence of an evaporation-induced temperature change is its effect on the surface tension coefficient of the droplet --- since evaporation occurs preferentially near the contact line of a droplet \cite{Deegan_1997}, not uniformly over the surface, the temperature-induced changes in surface tension coefficient would also lead to Marangoni flows, with their associated stresses. (We assume that other sources of Marangoni stress, e.g., surfactant concentration gradients, are negligible \cite{ristenpart2007influence}.) Accordingly $\gamma(T)= \gamma(T_0) + \beta (T-T_0) + O(T-T_0)^2$, and the typical change in $\gamma$ caused by temperature changes is
$$\mathrm{Ma}=\frac{|\Delta\gamma|}{\gamma}\sim|\beta \Delta T/\gamma|\approx 0.05.$$
Let $H$ be the height of the droplet. Marangoni effects are therefore expected to induce a fluid flow of velocity $u_M$ such that $\mu u_M/H\sim \gamma \mathrm{Ma}/\ell_c$, so that the capillary number associated with the surface tension gradient is:
$$\mathrm{Ca_{Ma}}=\frac{\mu u_M}{\gamma}=\frac{\mathrm{Ma} H}{\ell_c}\lesssim1.$$ This capillary number is larger than the standard capillary number (so that the Marangoni effect is larger than the viscous stress induced by the evaporative flow) but is still small, so that the dominant physics in determining the droplet shape remains the bare interfacial tension.

\section{Mathematical model: derivation and numerical approximation \label{sec:Theory}}
In this section we give  details of the derivation and implementation of the gravito-elasto-capillary model for the shape evolution of the droplets during evaporation.
We distinguish two main phases during the evaporation of HFBI droplets:
\begin{enumerate}
\item \emph{Before the formation of the elastic, HFBI film}. In this phase the shape of the drop is dominated by the interplay of gravity and capillarity subject to evaporation with a constant contact area. In this phase, the shape of the drop is given by the classical Young--Laplace equation. As water evaporates, the HFBI concentration increases, though in this phase it is not sufficient to trigger the formation of an elastic film at the air--water interface.

\item \emph{After the formation of the HFBI film.} When the concentration of the hydrophobin molecules on the free surface is sufficient, HFBI self-organizes to form a solid film. In this phase, the shape of the drop is dictated by the elasticity of the membrane, the interfacial tension and the gravity acting on the drop.
\end{enumerate}

In all phases, we neglect concentration-dependent variation of the interfacial tension, though this is also possible.

\subsection{Phase 1: Shape determined by gravity and capillarity}
\label{sec:before_flattening}
In this phase we use  the theoretical framework proposed in \cite{boucher1980capillary} and expanded in \cite{wong2017non}. 
 In particular, we assume axisymmetry and we fix the center of a Cartesian frame $(x,\,y,\,z)$ at the apex of the drop. Let $(r(t,\,S), z(t,\,S))$ be the curve describing the shape of the droplet, where $r$ is the radial distance from the $z$ axis and $S$ is the arclength of the curve.  We denote by $\phi(t, S)$ the local tangent angle, so that
\begin{equation}
\label{eq:r'z'}
\left\{
\begin{aligned}
&\frac{\d r}{\d S}=  \cos \phi;\\
&\frac{\d z}{\d S}=  \sin \phi;\\
\end{aligned}
\right.
\end{equation}
Using these variables, we can express the total curvature of the surface as \cite{boucher1980capillary}
\[
\kappa =\pm\left( \frac{\d\phi}{\d S}+\frac{\sin \phi} {r}\right),
\]
the sign in front of the parenthesis is positive if we are considering a sessile droplet (the droplet lies below the curve $(r,\,z)$), while it is negative in the case of a pendant droplet (the droplet lies above the curve $(r,\,z)$).

Due to the hydrostatic pressure profile within the droplet, the pressure at a position $(r,z)$ can be written
\[
p = p_T - \rho g z = \pm\frac{\gamma}{V_i^{1/3}}\alpha - \rho g z
\]
 where $p_T$ is the pressure at $z=0$ and $\alpha=\pm V_i^{1/3} p_T/\gamma$ is dimensionless, with a positive or a negative sign depending on whether the droplet is sessile or pendant, respectively. Thus, we can write the Young--Laplace equation as
\begin{equation}
\label{eq:YL}
\rho g z - p_T= \pm\gamma\left(\frac{\d\phi}{\d S}+\frac{\sin \phi} {r}\right).
\end{equation} 

Using $V_i^{1/3}$ as the characteristic length scale of the problem, we introduce the following dimensionless quantities
\begin{equation}
\label{eq:non_dim_quant_1}
\tilde{r} = \frac{r}{V_i^{1/3}},\qquad
\tilde{z} = \frac{z}{V_i^{1/3}},\qquad
\tilde{S} = \frac{S}{V_i^{1/3}}.
\end{equation}
In the following, we drop the tildes from all the dimensionless variables for notational compactness. We can rewrite the  set of ODEs \eqref{eq:r'z'}-\eqref{eq:YL} in the following dimensionless form
\begin{equation}
\label{eq:non_dim_sis}
\left\{
\begin{aligned}
&\frac{\d r}{\d S} = \cos\phi,\\
&\frac{\d z}{\d S}=\sin\phi,\\
&\frac{\d\phi}{\d S}= \mathrm{Bo}\,z -  \alpha -\frac{\sin\phi}{ r}.
\end{aligned}
\right.
\end{equation}

To compute the shape of a droplet having a volume $V(t)$, we use the following initial conditions:
\begin{equation}
\label{eq:BC}
\left\{
\begin{aligned}
&r(t,\,0)=0,\\
&z(t,\,0)=0,\\
&\phi(t,\,0)=0.
\end{aligned}
\right.
\end{equation}
The system \eqref{eq:non_dim_sis} is integrated using the software \textsc{Mathematica} iteratively changing the value of $\alpha$. In particular, we solve the system \eqref{eq:non_dim_sis} until $\phi=-\pi$. Then, we identify the value(s) $S_\text{end}$ such that
\begin{equation}
\label{eq:target_r}
r(t,\,S_\text{end})=r_c,
\end{equation} 
where $r_c$ is the (dimensionless) radius of the contact surface between the drop and the substrate, which is measured from the experiments. We iterate this process until the volume of the droplet $V$ coincides with the target one $V(t)$.

We observe that the boundary conditions \eqref{eq:BC} are not suitable  for use in the numerical procedure because of the singularity of the Young--Laplace equation \eqref{eq:YL} when $r=0$. To circumvent this difficulty, we exploit the following series expansion:
\begin{equation}
\label{eq:series_no_elas}
\left\{
\begin{aligned}
&r(t,\,\varepsilon_0)=\varepsilon_0 r_1 + o(\varepsilon_0),\\
&z(t,\,\varepsilon_0)=\varepsilon_0 z_1 + o(\varepsilon_0),\\
&\phi(t,\,\varepsilon_0)=\varepsilon_0 \phi_1 + o(\varepsilon_0).\\
\end{aligned}
\right.
\end{equation}

Plugging \eqref{eq:series_no_elas} into \eqref{eq:non_dim_sis}, we find
\[
\left\{
\begin{aligned}
&r_1 = 1,\\
&z_1 = 0,\\
&\phi_1 = -\frac{\alpha}{2},\\
\end{aligned}
\right.
\]
so that, in place of \eqref{eq:BC}, we neglect the remainder in \eqref{eq:series_no_elas} and we use these expressions as initial conditions in the numerical algorithm, choosing a suitable, small $\varepsilon_0$.

\subsection{Phase 2: Shape determined by gravity, capillarity and elasticity}

Following \cite{Knoche_2013}, we can rewrite  Eq.~(3) of the main paper as a system of first order ordinary differential equations. The principal curvatures $\kappa_s$ and $\kappa_\theta$ can be expressed as \cite{Libai_1998,Knoche_2011}
\begin{equation}
\label{eq:kappas}
\kappa_s = \pm\frac{\d\phi}{\d s},\qquad\kappa_\theta =\pm \frac{\sin\phi}{r},
\end{equation}
where again we take a plus or a minus if we are considering a sessile or a pendant droplet, respectively.

Recalling that the constitutive equations for the meridional and hoop stresses are
\begin{equation}
\label{eq:tensions}
\left\{
\begin{aligned}
&\tau_s = \frac{E}{1-\nu^2} \frac{1}{\lambda_\theta}\left[(\lambda_s-1)+\nu(\lambda_\theta-1)\right]+\gamma,\\
&\tau_\theta = \frac{E}{1-\nu^2} \frac{1}{\lambda_s}\left[(\lambda_\theta-1)+\nu(\lambda_s-1)\right]+\gamma,\\
\end{aligned}
\right.
\end{equation}
Eq.~(3) of the main paper becomes
\begin{equation}
\label{eq:membrane_dimensional}
\left\{
\begin{aligned}
&\frac{\d r}{\d s_0} = \lambda_s \cos\phi,\\
&\frac{\d z}{\d s_0}=\lambda_s \sin\phi,\\
&\frac{\d\phi}{\d s_0}= \frac{\lambda_s}{\tau_s}\left(\pm(\rho g z - p_T) -\frac{\sin\phi}{r}\tau_\theta\right),\\
&\frac{\d\tau_s}{\d s_0}=\lambda_s\frac{\tau_\theta-\tau_s}{r}\cos\phi,
\end{aligned}
\right.
\end{equation}
where $\kappa_\theta,\,\tau_\theta$ and $\lambda_s$ can be obtained through \eqref{eq:tensions}, \eqref{eq:kappas} and the relationships
\begin{equation}
\lambda_s = \frac{\d s}{\d s_0},\qquad \lambda_\theta = \frac{r}{r_0}.
\label{eqn:stretches}
\end{equation}

As done for the drop without the elastic membrane, the system \eqref{eq:membrane_dimensional} is non-dimensionalized with respect to the characteristic length $V_i^{1/3}$ and the surface tension $\gamma$. We introduce the dimensionless functions \eqref{eq:non_dim_quant_1} and
\begin{equation}
\label{eq:non_dim_quant_2}
\tilde{\tau}_s = \frac{\tau_s}{\gamma},\qquad
\tilde{\tau}_\theta = \frac{\tau_\theta}{\gamma}.
\end{equation}
In addition to the parameters already discussed, this non-dimensionalization introduces a dimensionless interfacial modulus
$$\tE=E/\gamma.$$

For notational convenience, we drop the tildes to denote the dimensionless counterparts of the dimensional functions. The system \eqref{eq:membrane_dimensional} in dimensionless form becomes
\begin{equation}
\label{eq:membrane_non_dimensional}
\left\{
\begin{aligned}
&\frac{\d r}{\d s_0} = \lambda_s \cos\phi,\\
&\frac{\d z}{\d s_0}=\lambda_s \sin\phi,\\
&\frac{\d\phi}{\d s_0}= \frac{\lambda_s}{\tau_s}\left(\mathrm{Bo}\,z - \alpha -\frac{\sin\phi}{r}\tau_\theta\right),\\
&\frac{\d\tau_s}{\d s_0}=\lambda_s\frac{\tau_\theta-\tau_s}{r}\cos\phi,
\end{aligned}
\right.
\end{equation}
which is complemented by the initial conditions
\begin{equation}
\label{eq:BC_membr}
\left\{
\begin{aligned}
&r(t,\,0)=0,\\
&z(t,\,0)=0,\\
&\phi(t,\,0)=0,\\
&\tau_s(t,\,0)=\tau_{s0}.
\end{aligned}
\right.
\end{equation}
Finally, the membrane is assumed be constrained so that it is pinned at the contact line, namely
\begin{equation}
\label{eq:r_c_elas}
r(t,\,L)=r_c,
\end{equation}
analogously with \eqref{eq:target_r}.

The system is solved by fixing the value of $\tau_s$ and by performing a shooting method until the boundary condition \eqref{eq:r_c_elas} is satisfied. However, as before, the system is singular at $s=0$ since $r$ vanishes. Thus, we perform the following series expansion:
\begin{equation}
\label{eq:series_elas}
\left\{
\begin{aligned}
&r(t,\,\varepsilon)=\varepsilon r_1 +\varepsilon^2 r_2+ o(\varepsilon^2),\\
&z(t,\,\varepsilon)=\varepsilon z_0 + \varepsilon^2 z_2 + o(\varepsilon^2),\\
&\phi(t,\,\varepsilon)=\varepsilon \phi_1 +\varepsilon^2 \phi_2+ o(\varepsilon^2),\\
&\tau_s(t,\,\varepsilon)=\tau_{s0}+\varepsilon\tau_{s1}+\varepsilon^2 \tau_{s2} + o(\varepsilon^2),\\
\end{aligned}
\right.
\end{equation}
\subsubsection{Leading order expansion}
Substituting the expansion \eqref{eq:series_elas} into \eqref{eq:membrane_non_dimensional}, we find at  leading order that
\begin{equation}
\label{eq:leading_order} 
\left\{
\begin{aligned}
&r_1 = \frac{\tE}{(\nu -1) (\tau_{s0}-1)+\tE},\\
&z_1 = 0,\\
&\phi_1 = -\frac{\alpha  r_1}{2 \tau_{s0}},\\
&\tau_{s1} = 0.
\end{aligned}
\right.
\end{equation}
Thus, in place of \eqref{eq:BC_membr}, we neglect the remainder in \eqref{eq:series_elas} and we use these expressions as initial conditions in the numerical algorithm, choosing a suitable, small $\varepsilon$.

\subsubsection{Higher order terms}
Similarly to \eqref{eq:leading_order}, we can perform a series expansion of \eqref{eq:membrane_non_dimensional} at higher orders. In particular, second order terms in \eqref{eq:series_elas} read
\begin{equation}
    \label{eq:second_order}
\left\{
\begin{aligned}
&r_2 = 0,\\
&z_2 =-\frac{\alpha r_1^2}{4 \tau_{s0}},\\
&\phi_2 =0,\\
&\tau_{s2}=-\frac{(\tau_{s0}-1) (-\nu +2r_1-1) \left(4 \frac{\d^3r_0}{\d s_0^3}(0) \tau_{s0}^2+\alpha ^2r_1^2\right)}{64 (r_1-1)r_1 \tau_{s0}^2},
\end{aligned}
\right.
\end{equation}
\noindent where an expansion up to the third order in $r$ and $\phi$ is needed to compute $\tau_{s2}$.
The quantity $\d^3r_0/\d s_0^3(0)$ can be computed by expanding \eqref{eq:series_no_elas} up to the third order in $\epsilon_0$, obtaining
\[
\frac{\d^3 r_0}{\d s_0^3}(0)=-\frac{\alpha_0 ^2}{4},
\]
where $\alpha_0$ is the value of $\alpha$ in the initial configuration of the membrane.
For later convenience, we observe that if we similarly expand the hoop tension
\[
\tau_\theta(t,\,\epsilon) = \tau_{\theta0}+ \epsilon\tau_{\theta1}+\epsilon^2\tau_{\theta2}+o(\epsilon^2),
\]
a direct computation shows that
\begin{equation}
\label{eq:tau_theta}
\tau_{\theta0}=\tau_{s0},\qquad\tau_{\theta1} = 0,\qquad \tau_{\theta2} = 3\tau_{s2}.
\end{equation}

We remark that, as the meridional tension $\tau_s$ in $s=0$  goes to zero, some terms of the series expansion become singular. To avoid this issue, we first assume that $\tau_{s0}=0$ and then we perform the expansion \eqref{eq:series_elas}.
Specifically, we also get that $\alpha$ (i.e.~the dimensionless pressure at the apex of the droplet) is zero. An explicit expansion up to the second order is given by
\[
\sistema{
&r(t,\,\epsilon) =\epsilon r_1+o(\epsilon^2)= \frac{ \tE}{-\nu +\tE+1}\varepsilon + o(\varepsilon^2),\\
&z(t,\,\epsilon) = \epsilon^2 z_2 + o(\epsilon^2) = \pm\frac{ r_1 \sqrt{8 \text{Bo} (r_1-1) r_1^3+\alpha_0^2 (-\nu +2 r_1-1)}}{4 \sqrt{-\nu +2 r_1-1}}\epsilon^2 + o(\epsilon^2)\\
&\phi(t,\,\epsilon) = \frac{2 (\varepsilon^2z_2)}{(\varepsilon r_1)}+o(\varepsilon^2),\\
&\tau_s(t,\,\epsilon) =\frac{\text{Bo}}{8}(\varepsilon r_1)^2 + o(\varepsilon^2).
}
\]

Interestingly, the convexity of $\tau_s$ in $s_0=0$ depends on the the sign of $\mathrm{Bo}$: 
\begin{itemize}
    \item If $\mathrm{Bo}$ is positive, then $\tau_s$ is convex and has a local minimum at $s=0$; as a result, the apex of the sessile droplet is the nucleation point of buckling.
    
    \item If $\mathrm{Bo}$ is negative, then $\tau_s$ exhibits a local maximum at $s=0$ and the tension must first become negative for $s>0$, away from the apex. (The membrane must have already buckled in a neighbourhood of $s=0$ when the tension first becomes zero at the apex.)
    
\end{itemize}

We note also that, since $\tau_{\theta2}$ is three times $\tau_{s2}$ as shown in \eqref{eq:tau_theta}, then the hoop stress decreases faster than the meridional stress, even when $\tau_{s0}\neq0$. Thus, we expect that when compression occurs first away from the drop apex (i.e.~when $\mathrm{Bo}<0$), the first component of the stress to become zero as the droplet evaporates is $\tau_\theta$. As a result,  radial wrinkles should appear when $\mathrm{Bo}<0$ while for $\mathrm{Bo}>0$ bi-directional crumpling should occur (since $\tau_\theta$ and $\tau_s$ vanish simultaneously at the apex).

As a result, $\mathrm{Bo}=0$ is a sharp transition point, where the nature of the membrane buckling changes dramatically. In the following, we detail the numerical algorithms used to explore the two cases, that is $\mathrm{Bo}>0$ for sessile and $\mathrm{Bo}<0$ for pendant droplets.

\subsubsection{Flattening in sessile droplets}
For sessile droplets, we find that the hoop and radial stresses vanish at the centre of the droplet. As discussed in the main paper, this leads to a flat region at the centre of the droplet: since both $\tau_s$ and $\tau_\theta$ vanish, locally the HFBI film cannot sustain a pressure difference across the membrane, it must remain flat, and, further, $\alpha=0$.
While $\tau_s$ must be continuous at $s_0=s_f$ to ensure the global balance of the forces acting on the whole flat region, in principle $\tau_\theta$ can be discontinuous at $s_0=s_f$. However, since 
\[
\lim_{s_0\rightarrow s_f^+}\phi(s_0)
\]
is, in general, different from zero, we observe that from the third equation of \eqref{eq:membrane_non_dimensional} also $\tau_\theta$ must vanish as $s_0 \rightarrow s_f^+$.
Using \eqref{eq:tensions}, we get
\[
\lim_{s_0\rightarrow s_f^+}\lambda_s=\lim_{s_0\rightarrow s_f^+}\lambda_\theta=\frac{\tE}{\tE+1-\nu}.
\]
From the expression of the stretches \eqref{eqn:stretches} we get
\[
r(t,\,s_f) = \frac{\tE}{\tE+1-\nu}r_0(t,\,s_f).
\]

Summing up, fixing the value $s_f$, we need to integrate the system \eqref{eq:membrane_non_dimensional} from $s_0=s_f$ up to $s_0=L$ using the initial conditions
\[
\left\{
\begin{aligned}
&r(t,\,s_f)=\frac{\tE}{\tE+1-\nu}r_0(t,\,s_f),\\
&z(t,\,s_f)=0,\\
&\phi(t,\,s_f)=\phi_f,\\
&\tau_s(t,\,s_f)=0;
\end{aligned}
\right.
\]
Again, these initial conditions are unsuitable to be directly used numerically, since the fourth equation of \eqref{eq:membrane_non_dimensional} has a singularity when $\tau_s = 0$. To circumvent the problem, we exploit the following asymptotic expansion
\begin{equation}
\label{eq:expans}
\left\{
\begin{aligned}
&r(t,\,s_f)=\frac{\tE}{\tE+1-\nu}r_0(t,\,s_f) + \varepsilon_f r_1 + o(\varepsilon_f),\\
&z(t,\,s_f)= \varepsilon_f z_1 + o(\varepsilon_f),\\
&\phi(t,\,s_f)=\phi_f + \varepsilon_f \phi_1 + o(\varepsilon_f),\\
&\tau_s(t,\,s_f)=\varepsilon_f^2\tau_{s2} + o(\varepsilon_f^2).
\end{aligned}
\right.
\end{equation}
By substituting these expressions into \eqref{eq:membrane_non_dimensional} and  performing a Taylor series expansion, one obtains
\[
\left\{
\begin{aligned}
&r_1 =\frac{\tE \cos \phi_f}{\tE+1-\nu},\\
&z_1 =-\frac{\tE \sin \phi_f}{\tE+1-\nu},\\
&\phi_1 = \frac{\cot \phi_f \left(\frac{\d r_0}{\d s_0}(s_f) \left(2\frac{\d r_0}{\d s_0}(s_f)+\cos \phi_f\right)-r_0(s_f)\frac{\d^2 r_0}{\d s_0^2}(s_f)-3 \cos ^2\phi_f\right)}{r_0(s_f) \left(3 \cos \phi_f-2\frac{\d r_0}{\d s_0}(s_f)\right)},\\
&\tau_{s2} = \frac{\left((\tE-\nu )^2-1\right) \cos \phi_f \left(\cos \phi_f-\frac{\d r_0}{\d s_0}(s_f)\right)}{2 \tE r_0(s_f)^2},\\
\end{aligned}
\right.
\]
where the expression of $\phi_1$ can be obtained only if we take into account the terms $\varepsilon_f^2 r_2$ and $\varepsilon_f^2 z_2$ in the expansion of $r$ and $z$ \eqref{eq:expans}, even though we do not report explicitly their expressions. 

Summing up, the shape equations \eqref{eq:membrane_non_dimensional} are integrated numerically from $s_0=s_f+\varepsilon_f$ up to $s_0=L$, fixing $\alpha=0$ and by selecting a particular value of $s_f$. As initial conditions, we use \eqref{eq:expans} and perform a shooting method, changing the value of $\phi_f$ until the boundary condition \eqref{eq:r_c_elas} is satisfied.

\begin{figure}[t!]
    \centering
    \includegraphics[width=0.5\textwidth]{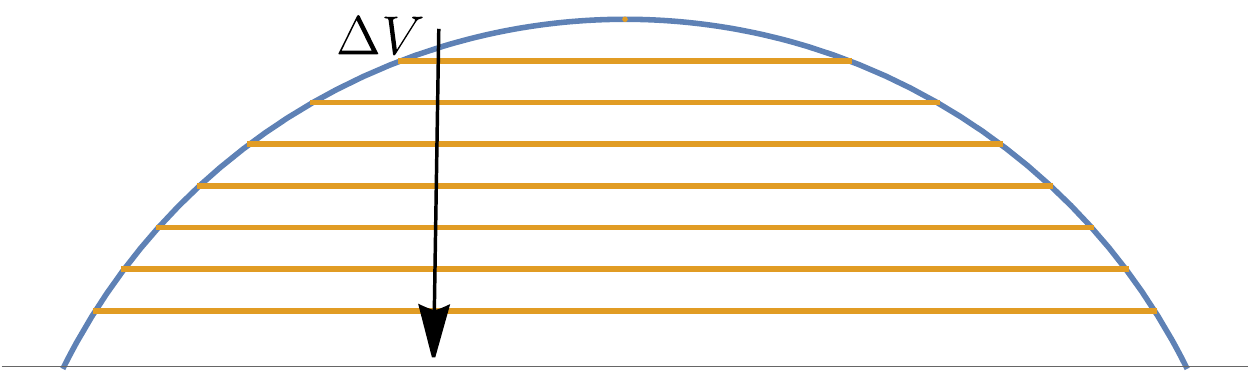}
    \caption{Representation of the ``spherical segment'' geometrical approximation, where the flattened droplet is modeled as a spherical segment. As water evaporates we remove a spherical cap from the apex of the droplet whose volume corresponds to the evaporated volume.}
    \label{fig:spherical_segment}
\end{figure}

\paragraph{Spherical segment approximation}
A simple geometrical model allows us to obtain a very close approximation of the diameter of the flattened region by approximating the droplet shape as a spherical segment in which the radius of curvature of the sphere and the contact line radius are both kept constant, as shown in FIG.~\ref{fig:spherical_segment}. 

This approximation may be justified by the high energetic cost of changing the Gaussian curvature in regions where the stress is significant --- it is energetically cheaper to extend the flat spot radius since the both components of the stress are close to zero here, by assumption.

Given the volume $V_f$ of the droplet at the onset of crumpling and the radius $r_c$ of the contact line, we can compute the height $h$ of the spherical cap using the following formula
\begin{equation}
\label{eq:Vf}
V_f = \frac{\pi h}{6}(3r_c^2 + h^2).
\end{equation}
By applying the Pythagorean theorem, we get
\[
R= \frac{h^2+r_c^2}{2 h}.
\]
Finally, we can find the evaporated volume $V_f-V$ for a given diameter $d$ of the flattened region by using again the formula for the volume of a spherical cap, as in \eqref{eq:Vf}, obtaining
\[
V_f-V = \frac{\pi}{12}   \left(2 R-\sqrt{4 R^2-d^2}\right) \left(\frac{1}{4} \left(2 R-\sqrt{4 R^2-d^2}\right)^2+\frac{3 d^2}{4}\right)
\] and hence
\begin{equation}
\frac{V_f-V}{V_f} =\frac{\pi}{12}   \left(\frac{2 R}{V_f^{1/3}}-\sqrt{\frac{4 R^2-d^2}{V_f^{2/3}}}\right) \left(\frac{1}{4V_f^{2/3}} \left(2 R-\sqrt{4 R^2-d^2}\right)^2+\frac{3 d^2}{4V_f^{2/3}}\right).    
\end{equation} 
The geometrical nature of this relationship suggests that the rescaled flat spot diameter, $d/V_f^{1/3}$, should be a function only of the rescaled volume change, $(V_f-V)/V_f$, as is observed experimentally. 

Moreover, and despite the crude nature of this simple approximation, we nevertheless find very good agreement with the experimental results, as shown in FIG.~2 of the main paper.

\subsubsection{Wrinkling in pendant droplets}
In contrast to sessile droplets, we find that in pendant droplets, the hoop stress vanishes \emph{before} the radial stress does. To model wrinkling in pendant droplets we therefore numerically solve \eqref{eq:membrane_non_dimensional} until either we reach $s_0=L$ or $\tau_\theta$ becomes zero at $s_a\in(0,\,L)$. In the latter case we switch to the far from threshold model developed in \cite{Knoche_2013}, maintaining $\tau_\theta=0$ in some wrinkled region. Since  $\tau_\theta=0$ in this region, we get
\begin{equation}
    \label{eq:lambdatheta_wrink}
\lambda_\theta = 1 - \frac{1-\nu^2}{\tE}\lambda_s - \nu (\lambda_s - 1),
\end{equation}
which is the actual stretch of the wrinkled surface, while the stretch of the axisymmetric pseudosurface is denoted by
\[
\widehat{\lambda}_\theta  = \frac{r}{r_0}.
\]

Substituting \eqref{eq:lambdatheta_wrink} into \eqref{eq:tensions} we get
\begin{equation}
    \label{eq:taus_wrink}
\tau_s =  \frac{\tE}{\lambda_\theta}\left(\lambda_s-1-\frac{\nu}{\tE}\lambda_s\right)+1,
\end{equation}
which represents the meridional strain in the wrinkled configuration. The meridional stress $\widehat{\tau}_s$on the pseudosurface can be computed from \eqref{eq:taus_wrink} as
\[
\widehat{\tau}_s = \frac{\lambda_\theta}{\widehat{\lambda}_\theta}\tau_s.
\]

Thus, on the pseudosurface the balance of forces becomes
\[
\frac{\d(r\widehat{\tau}_s)}{\d s} = 0,
\]
and therefore the system \eqref{eq:membrane_non_dimensional} can be rewritten as
\begin{equation}
\label{eq:wrinkled_non_dimensional}
\left\{
\begin{aligned}
&\frac{\d r}{\d s_0} = \lambda_s \cos\phi,\\
&\frac{\d z}{\d s_0}=\lambda_s \sin\phi,\\
&\frac{\d\phi}{\d s_0}= \frac{\lambda_s}{\tau_s}\left(\mathrm{Bo}\,z-\alpha\right),\\
&\frac{\d\widehat{\tau}_s}{\d s_0}=-\lambda_s\frac{\widehat{\tau}_s}{r}\cos\phi.
\end{aligned}
\right.
\end{equation}
The system is solved for $s_0\in[s_a,\,s_b]$, namely in the interval where
\[
\widehat{\lambda}_\theta<1 - \frac{1-\nu^2}{\tE}\lambda_s - \nu (\lambda_s - 1),
\]
while in $[s_b,\,L]$ we switch back to the membrane model \eqref{eq:membrane_non_dimensional}.

\begin{figure}[t!]
    \centering
    \includegraphics[width=0.5\textwidth]{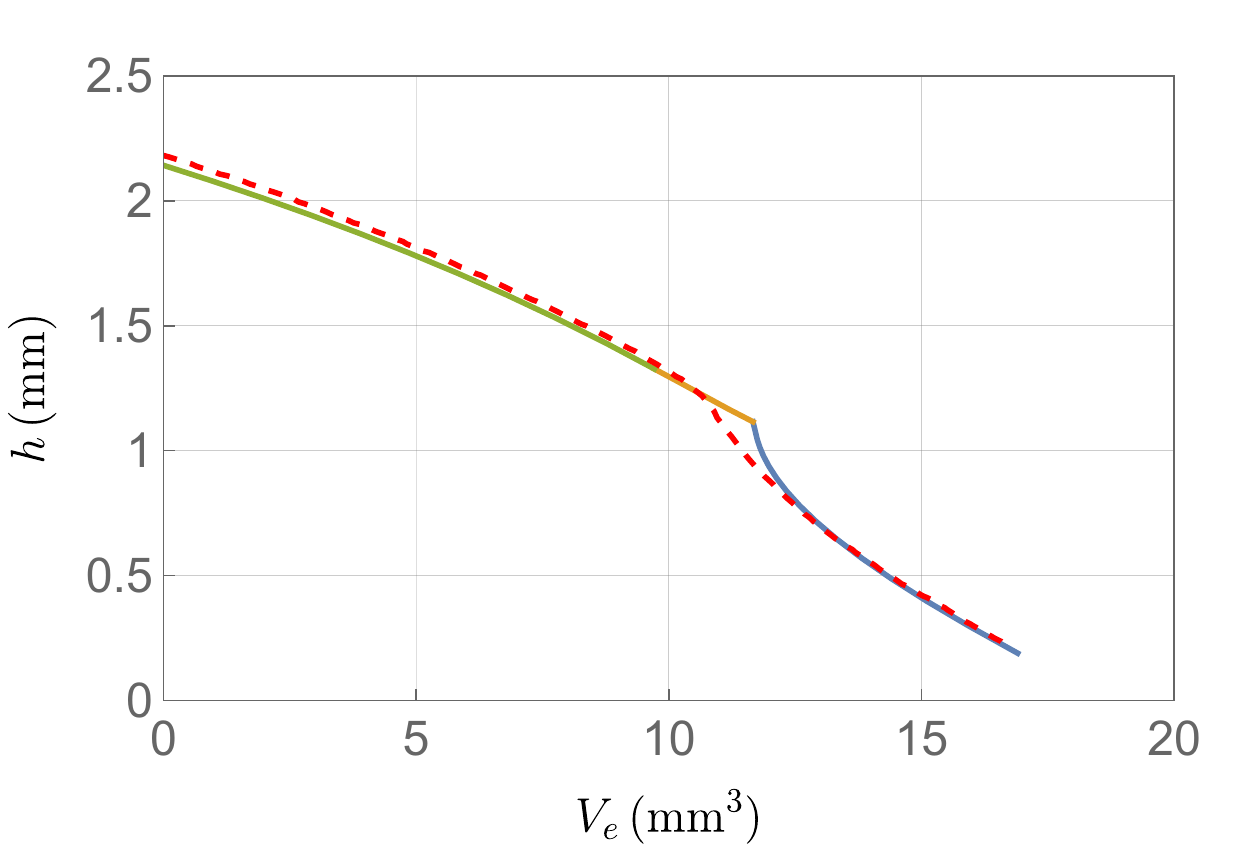}%
    \includegraphics[width=0.5\textwidth]{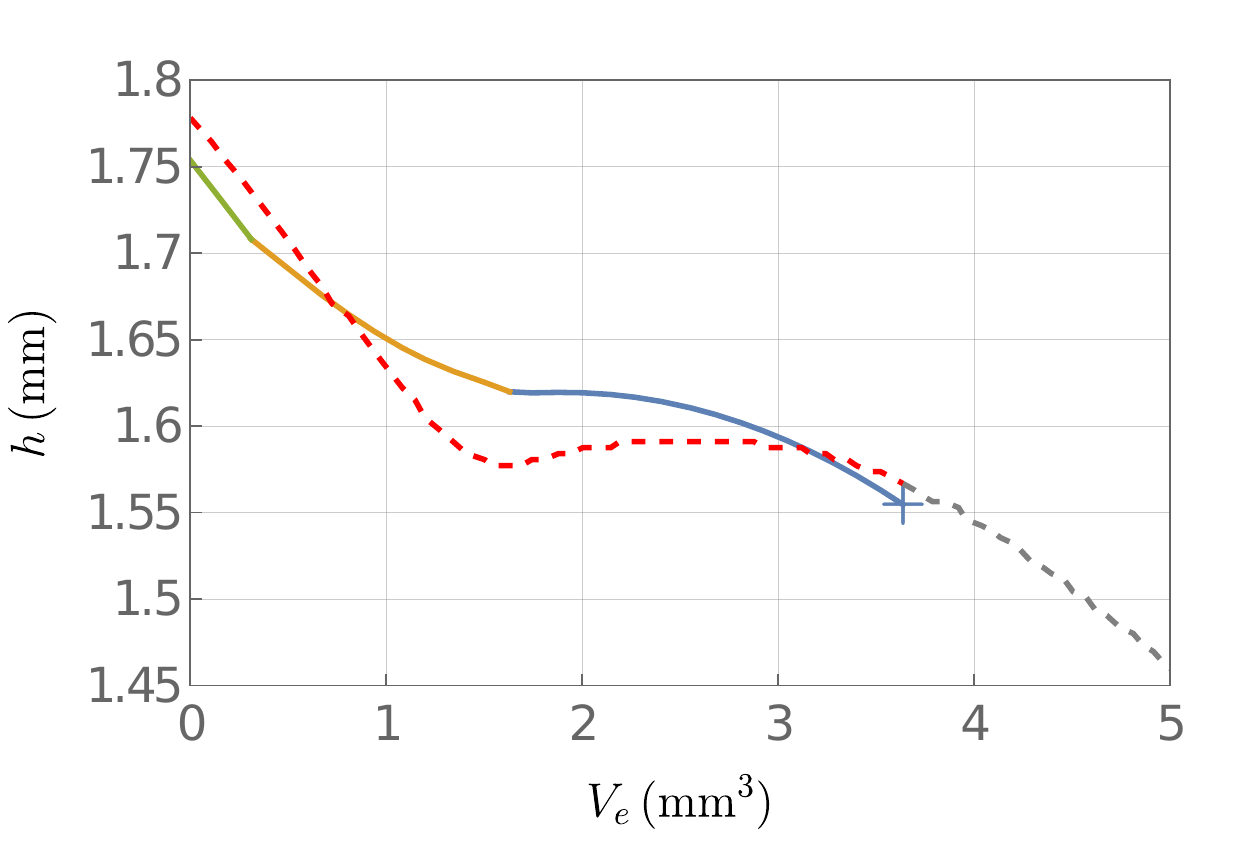}
    \caption{Height of the droplet $h$ as a function of the evaporated volume $V_e$ for  (left) a sessile droplet with initial volume $V_i = \SI{18.68}{\milli\meter^3}$ and initial concentration $C = \SI{2}{\micro\mole\per\liter}$ and (right) a pendant droplet with initial volume $V_i = \SI{7.77}{\milli\meter^3}$ and initial concentration $C = \SI{4}{\micro\mole\per\liter}$. The dashed line is extracted from the experimental data.  The experimental curve are plotted in gray at points corresponding to a depinned contact line. The solid curve represents the model prediction.  Specifically, the green curve corresponds to Phase 1, where the elastic membrane is not present, the orange curve corresponds to Phase 2 before wrinkling/crumpling, when the HFBI film is formed, and the blue curve indicates that crumples (for the sessile droplet) and wrinkles (in the pendant droplet) are present.  The point at which the model predicts the interpenetration of the membrane with the substrate is denoted `$+$' --- simulations cease at this point.}
    \label{fig:hvsVev}
\end{figure}

In FIG.~\ref{fig:hvsVev} we show the plots of the height of the droplet as a function of the evaporated volume for both sessile and a pendant droplets.
Note that, for pendant droplets, the shape predicted by the model interpenetrates the substrate; we indicate this point in FIG.~\ref{fig:hvsVev} with `$+$' and cease the simulation at this point. Interestingly, however, this point corresponds almost exactly with the point at which the contact line is observed to depin experimentally (indicated by the transition from the red to the gray dashed line in the right-hand plot of FIG.~\ref{fig:hvsVev}). In particular, the ``depinning'' of the contact line seems to be induced by the contact between the membrane and the substrate. Such an interaction is not modeled in this work and will be studied in a future paper.

\begin{figure}[t!]
    \centering
    \includegraphics[width=0.5\textwidth]{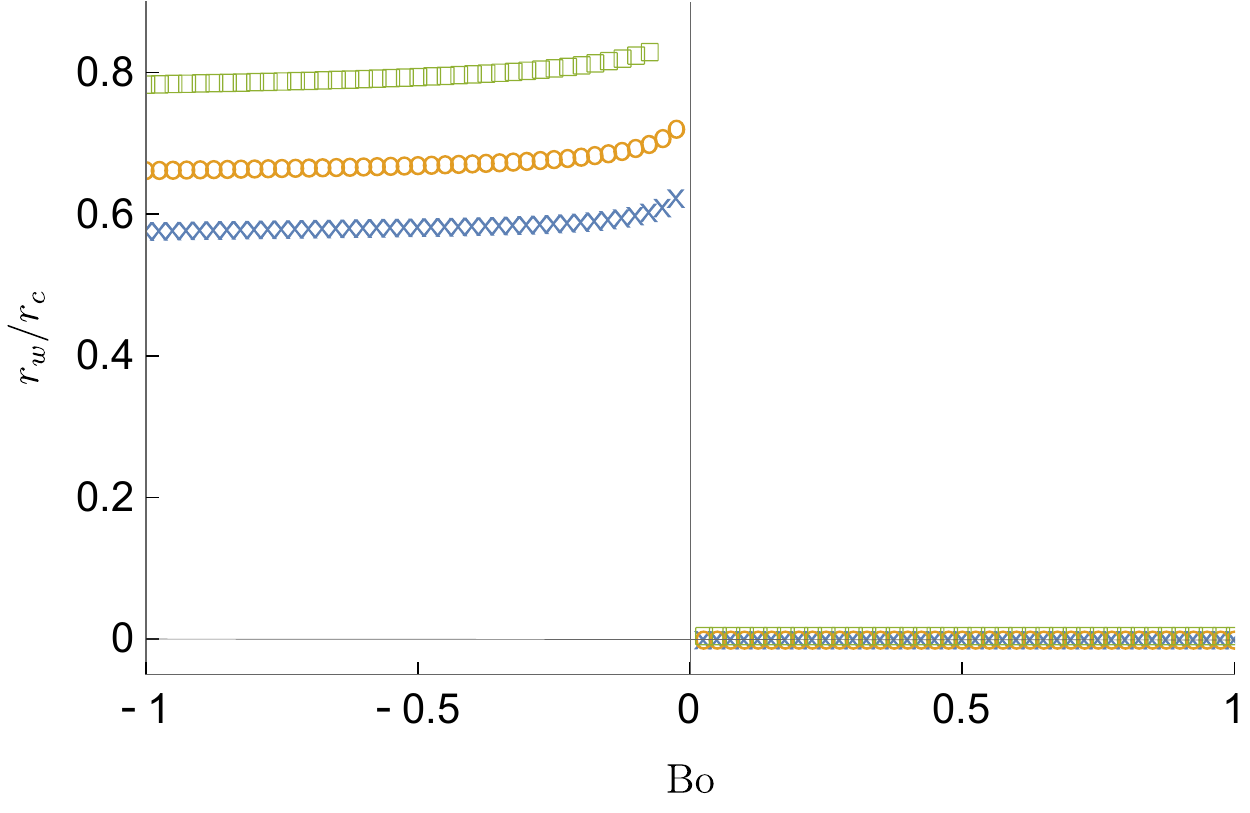}
    \caption{Plot of the normalized radius $r_w/r_c$ versus $\mathrm{Bo}$. Here, $r_w$ is the radial position where wrinkling/crumpling first appear. In the plots, $V_i = \SI{7.77}{\milli\meter^3}$, $C = \SI{4}{\micro\mole\per\liter}$, $\gamma = \SI{55}{\milli\newton\per\meter}$ while $E = 150,\,200,\,\SI{400}{\milli\newton\per\meter}$ (blue crosses, orange circles, and green squares, respectively). } 
    \label{fig:bovsrw}
\end{figure}

FIG.~\ref{fig:bovsrw} shows how the nucleation point of wrinkles for the pendant drop case (i.e.~the first point where $\tau_\theta$ becomes zero as water evaporates) varies with the two key dimensionless parameters: the ratio $\tE=E/\gamma$ and $\mathrm{Bo}<0$. We note that as $\tE$ increases or $\mathrm{Bo}$ approaches zero, the nucleation point moves closer to the contact line. This is noteworthy because it emphasizes the discontinuity in behavior that is observed around the zero Bond number limit: as Bond number $\mathrm{Bo}\nearrow0$ the nucleation point moves towards the contact line but for $\mathrm{Bo}>0$ the nucleation point lies precisely at the center of the droplet.

